\documentclass[journal=jctcce,manuscript=article]{achemso} 
\usepackage[T1]{fontenc} 
\usepackage[version=3]{mhchem} 
\usepackage[dvipsnames]{xcolor}
\usepackage{graphicx}
\usepackage{subcaption}
\usepackage{amssymb}
\usepackage{overpic}
\usepackage{soul}
\usepackage{bm}
\usepackage{comment}
\usepackage{titlesec}
\titleformat{\paragraph}
{\normalfont\normalsize\bfseries}{\theparagraph}{1em}{}
\titlespacing*{\paragraph}
{0pt}{3.25ex plus 1ex minus .2ex}{1.5ex plus .2ex}

\newif\ifhighlightchanges
\highlightchangesfalse

\definecolor{darkblue}{HTML}{003D6D}

\newcommand{\bR}{{\bm R}}
\newcommand{\br}{{\bm r}}
\newcommand{\bP}{{\bm P}}

\newcommand{\bL}{{\bm L}}

\newcommand{\bG}{{\bm G}}

\newcommand{\bX}{{\bm X}}
\newcommand{\bGamma}{{\bm \Gamma}}

\newcommand{\hH}{\hat H}

\newcommand{\hp}{\hat p}
\newcommand{\hx}{\hat x}
\newcommand{\hy}{\hat y}

\newcommand{\hbp}{{\hat{\bm p}}}
\newcommand{\hbP}{{\hat{\bm P}}}
\newcommand{\hbr}{{\hat{\bm r}}}

\newcommand{\bB}{{\bm B}}
\newcommand{\bA}{{\bm A}}

\newcommand{\mansia}{{44}}
\newcommand{\mansib}{{46}}
\newcommand{\mansic}{{49}}
\newcommand{\mansid}{{50}}
\newcommand{\mansie}{{101}}
\newcommand{\mansif}{{111}}

\newcommand{\bra}[1]{\left< #1\right|}
\newcommand{\ket}[1]{\left|#1 \right>}
\newcommand{\pp}[2]{\frac{\partial {#1}}{\partial {#2}}}
\newcommand{\hbm}[1]{\hat{\bm{#1}}}

\title{A Phase-Space Electronic Hamiltonian for Molecules in a Static Magnetic Field II: Quantum Chemistry Calculations with Gauge Invariant Atomic Orbitals}

\usepackage{geometry}
\author{Mansi Bhati}
\affiliation{Department of Chemistry, Princeton University, Princeton, NJ USA}
\author{Zhen Tao}
\affiliation{Department of Chemistry, Princeton University, Princeton, NJ USA}
\author{Xuezhi Bian}
\affiliation{Department of Chemistry, Princeton University, Princeton, NJ USA}
\author{Jonathan Rawlinson}
\affiliation{Department of Mathematics, Nottingham Trent University, Nottingham, UK}
\author{Robert Littlejohn}
\affiliation{Department of Physics, University of California, Berkeley, California 94720, USA}
\author{Joseph E. Subotnik}
\email{js8441@princeton.edu}
\affiliation{Department of Chemistry, Princeton University, Princeton, NJ USA}

\begin{document}

\maketitle

\begin{abstract}
In a companion paper, we have developed a phase-space  electronic structure theory of molecules in magnetic fields, whereby the electronic energy levels arise from diagonalizing a phase-space Hamiltonian $\hat H_{PS}(\bX,\bm{\Pi})$ that depends parametrically on nuclear position and momentum. The resulting eigenvalues are translationally invariant; moreover, if the magnetic field is in the $z-$direction, then the eigenvalues are also invariant to rotations around the $z-$direction.   However, like all Hamiltonians in a magnetic field, the theory has a gauge degree of freedom (corresponding to the position of the magnetic origin in the vector potential), and  requires  either $(i)$ formally, a complete set of electronic states or $(ii)$ in practice, gauge invariant atomic orbitals (GIAOs) in order to realize such translational and rotational invariance. Here we describe how to implement a phase-space electronic Hamiltonian using GIAOs within a practical electronic structure package (in our case, Q-Chem). We  further show that novel phenomena can be observed with finite $\bB-$fields, including minimum energy structures with  $\bm{\Pi}_{min} \ne 0$, indicating non-zero electronic motion in the ground-state.
\end{abstract}

\section{Introduction: Dynamics in Magnetic Fields}\label{intro}

The Born-Oppenheimer (BO) approximation is the bedrock  of quantum chemistry\cite{born1996dynamical}.  Based on the mass difference between the heavy nuclei and the light electrons, the BO approximation posits that one can construct  electronic stationary states with frozen nuclei.  For many processes of interest, the BO framework works quantitatively, as shown by vibrational spectroscopy\cite{bernathbook}.  However, in the presence of velocity dependent forces, such as in presence of magnetic fields, the BO approximation often fails\cite{cederbaum1988,cederbaum1991}.  The root cause of this failure is that, when nuclei move, such motion also induces changes in electronic motion during which time electrons should experience the impact of the magnetic field. However, within the BO approximation, there is no such feedback on the electronic degrees of freedom -- after all, electronic stationary states are calculated with frozen nuclei. Thus, at bottom, BO theory prescribes unequal magnetic shielding effects for electrons and nuclei\cite{Schmelcher_1988,yin1994magnetic}, and this imbalance has consequences. For instance and most famously, BO theory fails entirely to predict vibrational circular dichroism directly, which has necessitated going beyond  a BO formalism\cite{nafie1983SOS,Nafie1977BO,BUCKINGHAM1987,Stephens1985, duston2024phase}.

Now, within theoretical chemistry and physics, some of the limitations above can be treated by including a geometric potential in the Hamiltonian\cite{mead1979determination, berry1984quantal,kolodrubetz2017geometry}. In particular, there is now a growing list of theorists who run classical dynamics with Berry curvature forces.\cite{peters2021ab,Culpitt2022, peters2023berry, zhang2023tracking, Bian2023}  
Including the Berry curvature is clearly an improvement over standard BO dynamics, as the former allows one to recover momentum conservation without implicitly forcing the electronic momentum to be zero. That being said, including a Berry force is not a sinecure. After all, (i) the calculation of Berry curvature can be computationally expensive depending the level of theory. Although HF or DFT Berry forces can be constructed with a reasonable cost\cite{Culpitt2022},  excited state calculations would appear to be much more expensive [formally requiring the gradient of the derivative couplings].) (ii) Berry forces and derivative couplings are not well defined in the case of degeneracy, and the arbitrary choice of one adiabatic state introduces an arbitrary gauge.  (iii) Including Berry forces in the context of nonadiabatic electronic motion is quite difficult\cite{coraline_ehrenfest,miao:2019:fssh:complex,wu:jcp:2021:first_attempt_complex,bian:jctc:2022:first_attempt}, with no known successes as far as we are aware of; the Berry force diverges at a conical intersection. 
(iv) Perhaps most importantly, including a nuclear Berry force in a calculation improves the prediction of nuclear observables but yields no information about the combined effect of nuclear motion and a magnetic field on the {\em electronic} dynamics. For all these reasons, there is a strong motivation to develop new and practical approaches to magnetic-field chemistry.
In a companion paper\cite{bhati2024paper1} (henceforward, Paper I), we have now developed an alternative approach for going beyond standard BO theory. Specifically, we have formulated a phase-space electronic Hamiltonian, $\hH_{PS}(\bX,\bP)$, whereby the Schrodinger equation for the electronic stationary states is parameterized by both nuclear positions and momenta through so-called $\bGamma$ terms that are effectively electron translation factors(ETFs)\cite{fatehi_etf,schneiderman:1969:pr:etf} and electron rotation factors(ERFs)\cite{tian_erf,athavale:2023:erf}.
 As demonstrated in Paper I, this Hamiltonian has many desirable features:
\begin{itemize}
\item The Hamiltonian reduces to a  meaningful, field-free phase-space Hamiltonian that is known to recover accurate VCD spectra\cite{duston2024phase} and electronic momenta\cite{tao2024basis}.

\item
If one diagonalizes the Hamiltonian, the eigenenergies are invariant to translation and rotation. Moreover, if one propagates dynamics through Hamilton's equations along an eigenstate, the resulting dynamics will conserve the total pseudomomentum and total canonical angular momentum in the direction of the external magnetic field (see Paper I \cite{bhati2024paper1}).

\item The approach is exact for the hydrogen atom. 

\item 
The electronic Schrodinger equation can be solved very efficiently. In particular, the new terms (the so-called $\bGamma$ terms) in the electronic Hamiltonian are one-electron terms which yield minimal extra cost during diagonalization. 

\item Lastly, this magnetic Hamiltonian displays gauge invariance.  To make this attribute as clear as possible, note that mathematically, magnetic Hamiltonians are parametrized by a vector potential $\bA$, with kinetic energy
\begin{eqnarray}
\hat{H} = \frac{(\bP-\bA(\hbr))^2}{2M}    
\end{eqnarray}
The vector potential is related to the magnetic field by 
\begin{eqnarray}
    \bB&=&\nabla \times \bA(\hbr)
\end{eqnarray}
and is not uniquely defined;
one can add a scalar function of $\nabla f(\hbr)$  and retrieve the same magnetic field: 
\begin{eqnarray}
    \bA’(\hbr) &=& \bA(\hbr) + \nabla f(\hbr)\\
    \nabla \times \bA’(\hbr) &=& \nabla \times \bA(\hbr) + \nabla \times \nabla f(\hbr) =  \bB \label{eq:bchange}
\end{eqnarray}
In particular, for a set of charges in a uniform magnetic field,  the vector potential can be chosen as:
\begin{eqnarray}
    \bA’(\hbr) &=& \bA(\hbr) - \frac{1}{2} \bB \times \bG = \frac{1}{2} \bB \times (\hbr -\bG) \label{eq:stdB}
\end{eqnarray}
for an arbitrary origin $\bG$ without altering the magnetic field (i.e. $f(\hbr) = -1/2 (\bB \times \bG)\cdot \hbr$). Now, for a meaningful theory, the eigenstates of the electronic Hamiltonian must be invariant to the function $\bG$. Indeed, for the electronic phase-space Hamiltonian in Paper I\cite{bhati2024paper1}, this invariance holds. Namely, if one changes the vector potential from $\bA$ to $\bA'$,  an eigenstate changes from $\ket{\psi}$ to $\ket{\psi'}$:
\begin{eqnarray}
   \psi’(\br) = \psi(\br) \exp(-\frac{ie}{\hbar} f(\br) ) \label{eq:wfnchange} 
\end{eqnarray}
\end{itemize}

The present paper concerns this last topic, namely the question of gauge invariance. 
The invariance of all observables to gauge stems from the transformations in Eq. \ref{eq:wfnchange}, but this equality holds only when exactly diagonalizing the full electronic phase-space Hamiltonian.  Unfortunately, however, quantum chemistry calculations are rarely, if ever, performed in a complete basis; rather one almost always employs a much smaller, efficient, atomic orbital (AO) basis. In such an incomplete basis,  Eq. \ref{eq:wfnchange} will not hold and one finds that the energy of the system is not translationally invariant.\cite{pulay2007shielding,magnetictrans}.

The problem noted above has been discussed for many decades in the quantum chemistry world\cite{pople1957theory,Pulay1993} -- albeit within the standard BO framework. At present, the most common solution is to run a calculation using gauge-including atomic orbitals (GIAOs)\cite{london1937theorie, helgaker1991giaome, pulay2007shielding} which have been incredibly successful at calculating NMR shielding constants.\cite{jacspulay, helgaker1999ab,gauss1993effects, wong2023silico}  Conceptually, GIAOs dictate that we choose a different gauge for every basis function depending on the associated atom. 
By forcing each basis function to have its own atomic centered gauge (see below for the technical details), GIAOs ensure translational and rotational symmetry of all observables -- even if Eq. \ref{eq:wfnchange} still does not hold. As a practical matter, on the one hand,  GIAOs facilitate faster energy convergence, thereby enabling the use of smaller basis sets.\cite{jacspulay}
On the other hand,  GIAOs do introduce some 
complexity and extra cost to standard BO calculations by necessitating complex phases\cite{helgaker1991giaome,andrewtealegiao}.  That being said, for phase-space electronic Hamiltonian--which are always complex valued by definition--the extra computational cost of GIAOs should not be large.

With this background in mind, in what follows below, our goal is to demonstrate how GIAOs can be included in a phase-space electronic Hamiltonian formalism so as to ensure that we can recover meaningful beyond-BO electronic orbitals and energies for molecules (described by a truncated atomic orbital basis set) in magnetic field. An outline of this paper is as follows. In Sec. \ref{MD_GIAOs}, as a matter of background,  we give a brief overview of Born-Oppenheimer Dynamics using GIAOs. We discuss the various GIAO matrix elements of the terms in Hamiltonian, and matrix elements of different momentum vectors in Sec. \ref{GIAO_momentum}. We further show that using GIAOs leads to translationally (Sec. \ref{giao_energy_trans}) and rotationally (Sec. \ref{giao_energy_rot}) invariant energies. Next, in Sec. \ref{semi_class_ham_PS} we outline how to construct a phase-space Hamiltonian in an atomic basis, and in Sec. \ref{ETF-ERF} we give explicit expressions for the necessary ETFs and ERFs in such a GIAO basis. 
In Secs. \ref{gamma_trans} and \ref{gamma_rot}, 
we further show that including these ETF and ERF terms  leads to translationally  and rotationally invariant energies. Armed with the necessary matrix elements, in Sec. \ref{pm-am-conserve} we derive the equations of motion from a phase-space Hamiltonian within a GIAO basis, so that we can definitively prove that such a phase-space formalism conserves the total pseudomomentum and the total canonical angular momentum around the field. Finally, in Sec. \ref{results} we present several results where we use the phase-space formalism to calculate energies in varying magnetic fields and highlight new non-BO features. We discuss our findings and conclude in Sec. \ref{conclude}.  

\subsection{Notation}
Given the large number of matrices required for the present text, here we will outline our notation going forward:
\begin{itemize}
\item Lower case $\bm r$ denotes electronic coordinates, capital case $\bm X$ denotes nuclear coordinates; more generally, capital letters denote nuclei and small letters denote electrons.
    \item An ``e'' subscript denotes an electronic observable and a ``n'' subscript denotes a nuclear observable.
    \item The charge of an electron is denoted $-e$ (i.e. we fix $e >0$). The charge of nucleus I is $Q_Ie$.     
    \item $\left\{\alpha, \beta, \gamma, \delta, \eta \right\}$ index the $x,y,z$ Cartesian directions. 
    \item Bold indicates three-dimensional quantities, e.g. $\br$. 
    \item  ${I,J,...}$ indexes nuclear centers and ${i,j,...}$ index electron.
    \item $\bG$ denotes the gauge origin.
    \item $\bB$ denotes a magnetic field corresponding to vector potential $A(\bX)$.    
    \item Hats indicate operators (either nuclear or electronic), e.g. $\hat{H}$. 
    \item $u,v$ index adiabatic states
    \item  $\left\{\mu, \nu, \lambda, \sigma \right\}$ index atomic orbitals $\left\{\phi_{\mu},\phi_{\nu} ,\phi_\lambda, \phi_\sigma\right\}$ which are often shortened to $\left\{{\mu},{\nu} ,\lambda, \sigma\right\}$. Note that we will  always assume that \ul{orbital $\mu$ is attached to atom $M$ and orbital $\nu$ is attached to atom $N$,}  and when notationally convenient, we will  refer to one such orbital as $\ket{\phi^{(M)}_{\mu}}$ or just $\ket{\mu_M}$ for shorthand.
    \item A tilde over the wavefunction ($\tilde \phi_\nu$) and operators ($\tilde h_{\mu\nu}$) denotes GIAO dressed wavefunctions and operators.  
    \item A prime indicates that the matrix elements are translated, e.g. $h'_{\mu\nu}$.
    \item A bar on the basis functions indicates that the orbitals have been rotated, e.g. $|\bar{\mu}\rangle$. 
\end{itemize}

\section{Review of Born-Oppenheimer Dynamics in An Atomic Orbital Basis: GIAOs and Origin Independence}\label{MD_GIAOs}

To begin our discussion, let us review the standard theory of GIAOs and the question of origin independence within the context of BO theory. The classical Born-Oppenheimer Hamiltonian replaces the operator $\hbP$  with  a classical $\bP$ and is of the form:  
\begin{eqnarray}
    \hat{H}_{BO}(\bX,\bB,\bG) &=& \sum_I \frac{1}{2M_I}(\bP_I-eQ_I{\mathbf{A({X}_I)}} )^2  + \hat{H}_e(\bX,\bB,\bG) \label{eq:HBO}
\end{eqnarray}
where the electronic Hamiltonian depends parametrically  on the nuclear positions:
\begin{eqnarray}
     \hat{H}_e(\bX,\bB,\bG) &=& \frac{1}{2m_e}\sum_{i=1}^{N_{e}} (\hat{\mathbf{p}}_i + e\mathbf{{A}(\hat{r}}_i))^2 + \frac{1}{2}\sum_{i,j\neq i}^{N_{e}} \frac{e^2}{4\pi \epsilon_0 |\hbr_i-\hbr_j|}\nonumber\\&&- \sum_{i=1}^{N_{e}}\sum_{I=1}^{N_{n}} \frac{Q_I e^2}{4\pi \epsilon_0 |\hbr_i-\bX_I|} + \frac{1}{2}\sum_{I,J\neq I}^{N_{n}} \frac{Q_IQ_J e^2}{4\pi \epsilon_0 |\bX_I-\bX_J|} 
\end{eqnarray}
The BO potential energy $E_e(\bX)$ is the eigenenergy of the electronic Schrodinger equation:
\begin{eqnarray}
    \hat{H}_{e}\ket{\Phi(\bm{X})} &=& E_{e} \ket{\Phi(\bm{X})} 
\end{eqnarray}

In an atomic orbital basis, the energy $E_e$ is given by:
\begin{eqnarray}
\label{eq:BOuv}
    E_{e}(\bm X,\bm G, \bm B) &=&   \sum_{\mu\nu}{D}_{\nu\mu} {h}_{\mu\nu} + \sum_{\mu\nu\lambda\sigma} {d}_{\nu\mu\sigma\lambda}{g}_{\mu\nu\lambda\sigma} 
\end{eqnarray} 
Here, ${h}_{\mu\nu}$ and ${g}_{\mu\nu\lambda\sigma}$ are the matrix elements of one and two electron operators. ${D}_{\nu\mu}$ and ${d}_{\nu\mu\sigma\lambda}$ are the one and two electrons density matrices. 
For any finite basis, $E_e$ (and therefore $E_{BO}$) depends (erroneously) on the origin $\bG$, and the role of GIAOs is to remove this dependency in a sensible fashion.


\subsection{Hamiltonian with GIAOs}\label{GIAO_HBO} 
Let $\phi^{(M)}_\mu(\br)$ be an atomic orbital centered on atom $M$. In the presence of a magnetic field, the GIAO $\tilde \phi_\mu$ (localized on atom M) is defined to be\cite{helgaker1991giaome,pulay2007shielding,andrewtealegiao}: 
\begin{equation}
    \Tilde{\phi}^{(M)}_\nu(\bB,\br) = \exp{(-\frac{ie}{\hbar } \mathbf{A}_M\cdot \mathbf{\br})} \phi_\mu(\mathbf{\br})= \exp{\Big(-\frac{ie}{2\hbar}(\bB\times \bX_{MG}) \cdot \br\Big)} \phi^{(M)}_\nu(\br)\
\end{equation}
Here, the vector potential 
on atom $M$ is
$\mathbf{A}_M = \frac{1}{2} \bB \times {\bX_{MG}}$ where  $\mathbf{X_{MG}}$  is the  position of nucleus  $M$ relative to the gauge center $\bG$, 
\begin{eqnarray}
    \mathbf{X}_{MG} &=& \mathbf{X}_M-\mathbf{G}
\end{eqnarray}
For future reference, let us also define the vector potential $\mathbf{A}_{MN}$:
\begin{eqnarray}
    \mathbf{A}_{MN} = \frac{1}{2}\bB \times \bX_{MN} \;\;\; \text{where}\;\;\;
        \mathbf{X}_{MN} = \bX_M-\bX_N
\end{eqnarray}

Now, if one considers the matrix elements of the kinetic energy in a typical AO basis, the one-electron matrix element is given by:
\begin{eqnarray}
{h}_{\mu\nu}(\mathbf{B,X}) &=& \left< \mu_M \middle| \frac{[-i\hbar\bm{\nabla}+eA(\hbr)]^2}{2m_e}-\sum_K \frac{Z_K}{\bm{\hat r}_K} \middle| \nu_N \right>
\end{eqnarray}
Clearly, this matrix elements depends in a nontrivial fashion on the choice of origin, $\bG$. 
By contrast,   consider the result with GIAOs, where the one electron momentum operator can act both on the raw basis function and on the phase of the basis function. If we write $\hat{h} = \hat{T}  + \hat{V}$, where $\hat{T}$ is the electronic kinetic energy and $\hat{V}$ is the electronic potential energy, the kinetic energy operator in such a GIAO basis is given by:
\begin{align}
     \tilde{T}_{\mu \nu}(\bB,\bX) &=& \left< \mu_M \exp{\Big(-\frac{ie}{\hbar}\mathbf{A}_M \cdot \br\Big)}\middle| \frac{[-i \hbar\nabla + eA(\hbr)]^2}{2m_e}\middle| \exp{\Big(-\frac{ie}{\hbar}\mathbf{A}_N \cdot \br\Big)} \nu_N \right >
\end{align}
Using the commutator
\begin{eqnarray}
    [-i\hbar\nabla + eA(\hbr), \exp{(-\frac{ie}{\hbar} \mathbf{A}_N \cdot \br) }] &=& -eA(\hbr_N)\exp(-\frac{ie}{\hbar} \mathbf{A}_N \cdot \br),
\end{eqnarray}
we can move the  exponential phase of the GIAO to the left. Performing such an operation twice gives: 
\begin{eqnarray}
    \tilde{T}_{\mu\nu}(\bB,\bX) &=& \left< \mu_M\middle| \exp{\Big(\frac{ie}{2\hbar}(\bB\times \bX_{MN}) \cdot \br\Big)}\frac{{\hat{\bm\pi}}_N^2}{2m_e}\middle| \nu_N \right> \label{eq:hgiao}
\end{eqnarray}
Here, we  have defined the relative momentum  $ {\hat{\bm \pi}}_N$ as
\begin{equation}
\bm {\hat{\pi}}_N = -i\hbar\bm{\nabla} + \frac{e}{2}(\bB \times \hbr_N) \label{eq:Ncenter}
\end{equation}
where $\hbr_N = \hbr-\mathbf{X}_N$ is the position operator relative to nucleus $N$. 
Alternatively in Eq. \ref{eq:hgiao}, one can also differentiate the GIAO factor on the $\ket{\mu_M}$ basis function to give the kinetic energy in terms of the kinetic momentum  centered on atom $M$:
\begin{eqnarray} 
\tilde{T}_{\mu\nu}(\bB,\bX) &=& \left< \mu_M\middle| \frac{{\hat{\bm\pi}}_M^2}{2m_e}\exp{\Big(\frac{ie}{2\hbar}(\bB\times \bX_{MN}) \cdot \br\Big)}\middle| \nu_N \right>
\\
 {\hat{\bm{\pi}}}_M &=& -i\hbar\bm{\nabla} + \frac{e}{2}(\bB \times \hbr_M) \label{eq:Mcenter}
\end{eqnarray}
The end result is that, as Eqs. \ref{eq:Ncenter} and \ref{eq:Mcenter} make clear, the resulting integral is gauge independent (in so far as $\bG$ does not appear). Similarly, the overlap and the two-electron integrals do not depend on $\bG$ in a basis of GIAOs:
\begin{equation}
    \tilde{S}_{\mu\nu}(\mathbf{B,X}) = \left< \mu_M\middle| \exp{\Big(\frac{ie}{2\hbar}(\bB\times \bX_{MN}) \cdot \br\Big)}\middle| \nu_N \right> \label{eq:sgiao}
\end{equation}
\begin{equation}
    \tilde g_{\mu\nu \lambda \sigma}(\mathbf{B,X}) = \left< \mu_M \nu_N \middle| \exp{\Big(\frac{ie}{2\hbar} \bB\cdot (\mathbf{X}_{MP} \times \mathbf{ r_1} + \mathbf{X}_{NQ} \times \mathbf{ r_2})\Big)}\bm{\hat{r}}_{12}^{-1}\middle| \lambda_P \sigma_Q \right> \label{eq:ggiao}
\end{equation}
Above, we let orbitals $\lambda$ and $\sigma$ be attached to atoms $P$ and $Q$. Note that, when using GIAOs, all of the integrals in Eq. \ref{eq:hgiao} and  Eqs. \ref{eq:sgiao}-\ref{eq:ggiao} depend explicitly on the magnetic field. 

\subsection{GIAO matrix elements of momentum}\label{GIAO_momentum}
In the context of magnetic fields, the notion of momentum becomes more complicated. In particular, while the canonical momentum operator $\bP$ (i.e., the operator dual to $\bX$) remains well-defined, two new momentum operators appear: (i) the kinetic momentum \begin{eqnarray}
   \hat{\bm{\pi}}_e &=&\hbp_e + \frac{e}{2}(\bB \times (\hat{\br}-\bG))
\end{eqnarray}
(which represents mass times velocity) and (ii) the pseudomomentum 
\begin{eqnarray}
\hat{\bm{k}}_e &=& \hbp_e - \frac{e}{2} (\bB \times (\hbr-\bG))
\end{eqnarray}
which along with nuclear pseudomomentum is conserved during dynamics and it is also the sum of the derivative couplings (as shown in Paper I). Finally, in Paper I, we also defined a fourth pseudomomentum-like operator, which is the pseudomomentum relative to atomic position $I$. 
\begin{eqnarray}
    \hat{\bm{k}}^I_e &=& 
    {\hat{\bm \pi}_e}- \frac{e}{2}\bm{B}\times (\hbr-\bX_I)\\ 
    & = & {\hat{\bm p}}_e - \frac{e}{2}\bm{B}\times (\hbr-\bG)+  e\bm{B}\times (\bm{X}_I-\bG)\label{eq:kIdefn}
\end{eqnarray}
Let us now evaluate the explicit matrix elements for these different electronic momenta when dressed with GIAOs. 

\begin{enumerate}
    
\item The canonical momentum is:
\begin{eqnarray}
    \tilde{p}^\alpha_{\mu\nu} &=& \left< \mu_M \exp{(-\frac{ie}{\hbar}\mathbf{A}_M\cdot \br)}\middle| \hat{p}_e^\alpha \middle|\exp{(-\frac{ie}{\hbar}\mathbf{A}_N\cdot \br)}\nu_N \right> \\
    &=&-\left< \mu_M \middle|\exp{(\frac{ie}{\hbar}\mathbf{A}_{MN}\cdot \br)} \Big(\frac{e}{2}\bB\times ( \frac{\bX_{N}+\bX_{M}}{2}-\bG)\Big)_\alpha\middle| \nu_N \right> \nonumber \\
    && +\frac{1}{2}\Bigg[\left< \mu_M \middle|\exp{(\frac{ie}{\hbar}\mathbf{A}_{MN}\cdot \br)} \middle| \hat{p}_e^\alpha\nu_N \right> +\left< \hat{p}_e^\alpha \mu_M \middle|\exp{(\frac{ie}{\hbar}\mathbf{A}_{MN}\cdot \br)}\middle|\nu_N \right> \Bigg]\nonumber \\
\end{eqnarray}

\item The kinetic momentum is:
\begin{eqnarray}
    \tilde{\pi}^\alpha_{\mu\nu} &=& \left< \mu_M \exp{(-\frac{ie}{\hbar}\mathbf{A}_M\cdot \br)} \middle| \left( \hat{p}_e^\alpha + \frac{e}{2} \Big(\bB \times (\hbr-\bG)\Big)_\alpha \right) \middle| \exp{(-\frac{ie}{\hbar}\mathbf{A}_N\cdot \br)}\nu_N \right> \nonumber\\\\&=&\left< \mu_M \middle|\exp{(\frac{ie}{\hbar}\mathbf{A}_{MN}\cdot \br)}( \frac{e}{2}\bB\times (\hbr-\frac{\bX_{N}+\bX_{M}}{2}))_\alpha\middle| \nu_N \right> \nonumber \\
    && +\frac{1}{2}\Bigg[\left< \mu_M \middle|\exp{(\frac{ie}{\hbar}\mathbf{A}_{MN}\cdot \br)} \middle| \hat{p}_e^\alpha\nu_N \right> +\left< \hat{p}_e^\alpha \mu_M \middle| \exp{(\frac{ie}{\hbar}\mathbf{A}_{MN}\cdot \br)}\middle|\nu_N \right> \Bigg]
\end{eqnarray}

\item 
The pseudomomentum is:
\begin{eqnarray}
    \tilde{k}^\alpha_{\mu\nu} &=& \left< \mu_M \exp{(-\frac{ie}{\hbar}\mathbf{A}_M\cdot \br)} \middle| \left(\hat{p}_e^\alpha - \frac{e}{2} (\bB \times (\hbr-\bG))_\alpha\right) \middle|\exp{(-\frac{ie}{\hbar}\mathbf{A}_N\cdot \br)}\nu_N \right> \nonumber \\\\
    &=&-\left< \mu_M \middle|\exp{(\frac{ie}{\hbar}\mathbf{A}_{MN}\cdot \br)} \frac{e}{2}\Big(\bB\times (\hbr + \frac{\bX_{N}+\bX_{M}}{2}-2\bG)\Big)
    _\alpha \middle| \nu_N \right> \nonumber \\
    && +\frac{1}{2}\Bigg[\left< \mu_M \middle|\exp{(\frac{ie}{\hbar}\mathbf{A}_{MN}\cdot \br)} \middle| \hat{p}_e^\alpha\nu_N \right> +\left< \hat{p}_e^\alpha \mu _M \middle|\exp{(\frac{ie}{\hbar}\mathbf{A}_{MN}\cdot \br)} \nu_N \right> \Bigg]
\end{eqnarray}

\item The pseudomomentum with respect to atom $I$ is:
\begin{eqnarray}
    \tilde{k}^{I\alpha}_{\mu\nu} &=& \left< \mu_M \exp{(-\frac{ie}{\hbar}\mathbf{A}_M\cdot \br)}\middle| \left( \hat{\pi}_e^\alpha -e\Big(\bB\times(\hbr-\bX_I)\Big)_\alpha \right) \middle|\exp{(-\frac{ie}{\hbar}\mathbf{A}_N\cdot \br)}\nu_N \right> \nonumber\\\\&=&\left< \mu_M\middle|\exp{(\frac{ie}{\hbar}\mathbf{A}_{MN}\cdot \br)}\Big(\frac{e}{2}\bB\times (\hbr-\frac{\bX_{N}+\bX_{M}}{2})\Big)_\alpha\middle| \nu_N \right> \nonumber \\&&-\left< \mu_M\middle|\exp{(\frac{ie}{\hbar}\mathbf{A}_{MN}\cdot \br)}\Big( e\bB\times (\hbr-\bX_I)\Big)_\alpha\middle| \nu_N \right> \nonumber \\
    && +\frac{1}{2}\Bigg[\left< \mu_M \middle|\exp{(\frac{ie}{\hbar}\mathbf{A}_{MN}\cdot \br)} \middle| p_e^\alpha\nu_N \right> +\left< p_e^\alpha \mu_M \middle|\exp{(\frac{ie}{\hbar}\mathbf{A}_{MN}\cdot \br)}\middle|\nu_N \right> \Bigg]
\end{eqnarray}

\end{enumerate}
Note that, in the equations above, only the $\tilde{\bm{\pi}}_{\mu\nu}$ and $\tilde{\bm{k}}_{\mu\nu}^I$ matrix elements are independent of the gauge origin $\bG$, a fact which motivates building the $\bGamma$ matrix elements from  the $\tilde{\bm{k}}_{\mu\nu}^I$ matrix elements.

\subsection{The GIAO Energy is Translationally Invariant}\label{giao_energy_trans}
Thus far, we have focused on whether or not the relevant electronic structure matrix elements depend on $\bG$ explicitly, but a far more important question is how these matrices transform under rotations and translations, as these properties will determine how the relevant observables transform as well. For example, let us now demonstrate the well-known fact \cite{pulay2007shielding} that when using GIAOs,  the total energy (of any given BO state) becomes independent of gauge origin. To prove such a fact, note that:
\begin{eqnarray} 
E_{{GIAO}}^{BO}(\bX,\bm G, \bm B) & = & \sum_{\mu\nu}\tilde{D}_{\nu\mu} \tilde{h}_{\mu\nu} + \sum_{\mu\nu\lambda\sigma} \tilde{d}_{\nu\mu\sigma\lambda}\tilde{g}_{\mu\nu\lambda\sigma}
\end{eqnarray} 
Now, as shown in Sec. \ref{GIAO_HBO}, the magnetic origin $\bG$ does not appear in the expressions for matrix elements $\tilde{h}_{\mu\nu}$ and $\tilde{g}_{\mu\nu\lambda\sigma}$
(recall that a $\sim$ signifies a GIAO basis and see Eq. \ref{eq:hgiao} above for the  $\tilde{T}_{\mu\nu}$ expression). Thus, when solving for the one electron density matrix $\tilde{D}_{\nu\mu}$ and two electron density matrix $\tilde{d}_{\nu\mu\sigma\lambda}$, these matrix elements must also have no dependence on $\bG$. Nevertheless, the matrix elements $\tilde{h}_{\mu\nu}$ and $\tilde{g}_{\mu\nu\lambda\sigma}$ do depend on the physical origin; these matrix elements are not translationally invariant. Therefore, to demonstrate that $E_{GIAO}^{BO}$ is translationally invariant, we must work a bit harder and consider how the wavefunctions change with translation as well. To that end, let us begin by noting the simple identity, 
\begin{align}
&\exp\Big(-\frac{ie}{2\hbar} \left( \mathbf{B\times (X+\Delta)}\right)\cdot (\hat{\br}+\mathbf{\Delta})\Big)\nonumber  \\
&= \exp(-\frac{ie}{2\hbar} \mathbf{B\times X}\cdot \hat{\br})\exp(-\frac{ie}{2\hbar}\mathbf{B\times X}\cdot \mathbf{\Delta})\exp(-\frac{ie}{2\hbar}\mathbf{B\times \mathbf{\Delta}}\cdot \hat{\br})\label{eq:identity}
\end{align}
From this identity, it follows that translated one-electron matrix elements (denoted by a prime below) pick up an additional phase which depends on the vector between the atoms on which the basis functions are centered, 
\begin{eqnarray}
\label{translate_hgiao}
\tilde h_{\mu \nu}' = \tilde h_{\mu \nu}\mathbf{(X+\Delta)} = \exp{\Big(\frac{ie}{2\hbar}(\bB\times \bX_{MN}) \cdot \mathbf{\Delta}\Big)}\tilde h_{\mu \nu}\mathbf{(X)}
\end{eqnarray}
Similarly, the translated two electron matrix elements are given by,
\begin{eqnarray}
\tilde g_{\mu \nu \lambda \sigma}' = \tilde g_{\mu \nu \lambda \sigma}\mathbf{(X+\Delta)} = \exp{\Big(\frac{ie}{2\hbar}(\bB\times \bX_{MP}) \cdot \mathbf{\Delta}\Big)} \exp{\Big(\frac{ie}{2\hbar}(\bB\times \bX_{NQ}) \cdot \mathbf{\Delta}\Big)} \tilde g_{\mu \nu \lambda \sigma}\mathbf{(X)}   
\end{eqnarray}
Therefore, altogether it follows that the Hartree-Fock equation $\tilde{F}'C'=\tilde{S}'C'\epsilon'$  can be satisfied by choosing a new set of molecular orbital coefficients  with the same GIAO phases (which preserves the orbital energies $\epsilon' = \epsilon$):
\begin{eqnarray}
\tilde C'_{\mu i} &=& \exp{\Big(-\frac{ie}{2\hbar}(\bB\times \bX_{MG}) \cdot \mathbf{\Delta}\Big)} \tilde{C}_{\mu i}    
\end{eqnarray}
This choice of molecular orbitals leads to a  density matrix $ \bm \tilde  D' =   \bm\tilde C' \bm  \tilde  C'^{\dagger} $ that satisfies (under translation):
\begin{eqnarray}
    \tilde D_{\nu\mu }' &=& \exp{\Big(-\frac{ie}{2\hbar}(\bB\times \bX_{MN}) \cdot \mathbf{\Delta}\Big)}  \tilde D_{\nu\mu }
\end{eqnarray}
With this ansatz, the phases in the Hamiltonian matrix elements cancel with the phases in the density matrix, resulting in a translationally invariant energy:
\begin{align}
    E(\bX+\mathbf{\Delta}) = \sum_{\mu\nu}\tilde D'_{\nu\mu }\tilde h'_{\mu\nu} + \frac{1}{2}\sum_{\mu\nu\lambda\sigma} \tilde D'_{\nu\mu }\tilde D'_{\sigma\lambda} \tilde g'_{\mu\nu\lambda\sigma} = E(\bX) \label{eq:transEgiao}
\end{align}
As a side note, we mention that the kinetic momentum $\tilde{\bm{\pi}}_{\mu\nu}$ transforms in the same way as $\tilde{h}_{\mu\nu}$, i.e., 
\begin{eqnarray}
    \tilde{\bm{\pi}}_{\mu\nu}(\bX_M + \mathbf{\Delta}, \bX_N + \mathbf{\Delta}) &=& \exp{\Big(\frac{ie}{2\hbar}(\bB\times \bX_{MN}) \cdot \mathbf{\Delta}\Big)} \tilde{\bm{\pi}}_{\mu\nu}(\bX_M , \bX_N)\label{eq:piuv}
\end{eqnarray}
Thus, this observable is independent of the origin as well: $\left<\bm{\hat{\pi}}\right> = \left<\bm{\hat{\pi}}'\right>$.

\subsection{The GIAO Energy is Rotationally Invariant}\label{giao_energy_rot}
Although less appreciated, the energy of a given electronic BO state is also rotationally invariant if GIAOs are employed. To demonstrate such invariance, consider how the one and two electron matrix elements  transform when dressed with GIAOs. Let  $\bR$ be the operator that rotate around the z axis by amount $\bm{\delta}=(0,0,\delta)$:
\begin{eqnarray}\label{eq:rot_defn}
    \bm{R} = \exp(-\frac{i}{\hbar}\sum_\alpha \bm{L}^\alpha \delta_\alpha) \approx 1 - \frac{i}{\hbar} \sum_{\alpha} \bm{L}^{\alpha}\delta_{\alpha} = 1 - \frac{i}{\hbar} {L}^{z}\delta 
\end{eqnarray}
Now, to correctly rotate the matrix elements, we must to rotate both the molecule and the atomic basis functions on them. To that end, let us  define $\left|\bar{\nu}_N \right>$ as the  orbital we found if we rotate the atomic orbital $\left|\nu_N \right>$.  In Appendix \ref{rot_proof}, we show that, for GIAO dressed one electron and two electron interaction operators, the following is true:
\begin{eqnarray}
    \label{eq:h_rot}
    \tilde h_{\bar{\mu}\bar{\nu}}(\bm{R} \bm X) &=& \tilde h_{\mu\nu}(\bm X)\\
    \label{eq:pi_rot}       \tilde \pi_{\bar{\mu}\bar{\nu}\bar{\lambda}\bar{\sigma}}(\bm{R} \bm X) &=&\tilde \pi_{\mu\nu\lambda\sigma} (\bm X)
\end{eqnarray}

Given the fact that all of that matrix elements are rotationally invariant, (and therefore the density matrix will  be unchanged under rotation), it follows that the total energy must also be rotationally invariant.

\section{A Phase-Space Electronic Hamiltonian}\label{semi_class_ham_PS}

As reviewed above, the Born-Oppenheimer potential energy becomes independent of the origin $\bG$ if we use GIAOs. Let us now prove the equivalent statement for phase-space potential energy surfaces.
In the presence of a magnetic field, the form of phase-space Hamiltonian as proposed in the companion paper is as below:
\begin{align}
\hat{H}_{PS}(\bm X,\bP,\bm G, \bm B)=\sum_{I}\frac{1}{2M_I} \left( \bP_I - q^{\textit{eff}}_I(\bX)A(\bX_I)-i\hbar \hbm{\Gamma}_I(\bX) \right)^2+  \hH_{e}(\bm X,\bm G, \bm B)\label{eq:PSH}
\end{align}
where we have defined the screened effective charge $q^{\textit{eff}}$ as:
\begin{eqnarray}
    q^{\textit{eff}}_I(\bm{X}) &=& Z_Ie -\left< \Psi \middle|\hat{\Theta}_I(\hbr,\bm{X})\middle| \Psi \right>
\end{eqnarray}
with $\hat{\Theta}_I(\hbr,\bm{X})$ given by:
\begin{eqnarray}
    \hat{\Theta}_I(\hbr,\bm{X}) &=&\frac{ m_{I} e^{-|\hbr-\bm{X}_I|^2/\sigma_I^2}}{\sum_J m_{J}e^{-|\hbr-\bm{X}_J|^2/\sigma_J^2}}\label{eq:thetadef}
\end{eqnarray}
Here, $\sigma_I$, was parameterized using electronegativity of atom $I$ and $\Psi$ is the phase-space wavefunction defined at zero magnetic field. Here we will refer to {\em raw} nuclear kinetic momentum  with effective nuclear charges as:
\begin{eqnarray}
   \bm{\Pi}^{\textit{eff}}_I&=&\bP_I - \frac{1}{2} q^{\textit{eff}}_I(\bX)(\bB\times (\bX_I -\bG))\label{eq:pieffdefn}
\end{eqnarray}
while the phase-space kinetic momentum is defined as:
\begin{eqnarray}
\bm{\Pi}^{\textit{PS}}_I&=&\bP_I - \frac{1}{2} q^{\textit{eff}}_I(\bX)(\bB\times (\bX_I -\bG)) - i\hbar \hbm{\Gamma}_I(\bX)\label{eq:pikindefn} \\
& = & \bm{\Pi}^{\textit{eff}}_I
- i\hbar \hbm{\Gamma}_I(\bX)
\end{eqnarray}
The $\hat{\bm{\Gamma}}$ term is the sum of electron translation factors  and electron rotational factors  terms as appropriate for the presence of a magnetic field. In Paper I, we have argued that a meaningful form of $\hat{\bm{\Gamma}}$ term (satisfying the required properties listed in Sec. 3.2 of Paper I)
is:
\begin{eqnarray}
    \bm{\hat{\Gamma}}_I &=& \bm{\hat{\Gamma}}_I' + \bm{\hat{\Gamma}}_I''\label{eq:gammasum}\\
     \bm{\hat{\Gamma}}_I' &=& \frac{1}{2i\hbar}\left( \hat{\Theta}_I(\hbr){\hat{\bm{ k}}_I}+{\hat{\bm{ k}}_I}\hat{\Theta}_I(\hbr)\right)\label{eq:etf}\\
     \hat{\bm{\Gamma}}_I^{''} &=& \sum_{J}  \zeta_{IJ}\left(\bm{X}_I -\bm{X}^0_{J}\right)\times \left(\bm{K}_J^{-1}\frac{1}{2i\hbar}(\hbm{r}-\bm X_J)\times(\hat{\Theta}_J(\hbr) \hat{\bm{k}}_{J} + \hat{\bm{ k}}_{J}\hat{\Theta}_J(\hbr))\right)\label{eq:erf}
\end{eqnarray}
where the $\zeta_{IJ}$, $\bm{X}_J^0$, and $\bm{K}_J$ are defined as follows:
\begin{eqnarray}
      \zeta_{IJ} &=& e^{-|\bm{X}_I-\bm{X}_J|^2/2(\sigma_I+\sigma_J)^2}\label{eq:zeta}\\
    \bm{X}_{J}^0 &=&\frac{\sum_I  \zeta_{IJ}\bm{X}_I}{\sum_I  \zeta_{IJ}}\\
    \bm{K}_J &=&\sum_I  \zeta_{IJ}\left(\bm{X}_I\bm{X}_I^\top-\bm{X}_J^0\bm{X}_J^{0\top}-(\bm{X}_I^\top\bm{X}_I-\bm{X}_J^{0\top}\bm{X}_J^0)\mathcal{I}_3\right)
\end{eqnarray}
The $\hat{\bm{k}}_{I}$ in the above equations is the electronic pseudomomentum of an electron with respect to a nuclei $I$ (see Eq. \ref{eq:kIdefn} above):
\begin{eqnarray}
    \hat{\bm{k}}_I  = \bm{\hat{\pi}}_e -e\bm{B}\times (\bm{\hat{r}}-\bm{X}_I)\label{eq:pseudolike}
\end{eqnarray}
With these definitions, the energy of the phase-space Hamiltonian in Eq. \ref{eq:PSH} in a GIAO basis representation is as follows:
\begin{eqnarray}
\label{eq:pssh_e}
    E_{\rm PS}(\bm X,\bP,\bm G, \bm B) &=& \sum_I\frac{(\bm{\Pi}_{I\alpha}^{\textit{eff}})^2}{2M_I} + V_{PS}(\bX,\bP,\bm G, \bm B) \\\label{eq:std_e}
    V_{PS}(\bX,\bP,\bm G, \bm B) & = & \sum_{\mu\nu}\tilde{D}_{\nu\mu} \tilde{h}_{\mu\nu} + \sum_{\mu\nu\lambda\sigma} \tilde{d}_{\nu\mu\sigma\lambda}\tilde{g}_{\mu\nu\lambda\sigma} \nonumber \\ &&-i\hbar\sum_{\mu\nu I\alpha}\tilde{D}_{\nu\mu}\frac{\bm{\Pi}_{I\alpha}^{\textit{eff}}\cdot \tilde{\bm{\Gamma}}^{I\alpha}_{\mu\nu}}{M_{I}}- \hbar^2\sum_{\mu\nu I}\tilde D_{\nu \mu}\frac{(\tilde{\bm \Gamma}^2_I)_{\mu\nu}}{2M_I}
\end{eqnarray} 
In Eq. \ref{eq:pssh_e}, we have separated the nuclear kinetic energy from the phase-space potential energy, which is the usual electronic energy plus a nuclear-electronic coupling term. 

Now, as pointed out in Paper I, in designing $\bGamma'$ and $\bGamma''$ above (Eqs. \ref{eq:etf}-\ref{eq:erf}), there are several constraints imposed by the need for pseudomomentum conservation and angular momentum conservation. See Eqs. $\mansia$ and $\mansib$, and Eqs. $\mansic -\mansid$  of Paper I for the basis-free form of these constraints. In a GIAO basis, these constraints take the form:
\begin{eqnarray}
    \left< \tilde{\mu}_M \middle|\Big( -i\hbar\sum_{I}{{ \hat{\bm{\Gamma}}}}_{I} + \sum_I { \hat{\Theta}_I}{ \hat{\bm k}_I}\Big)\middle| \tilde{\nu}_N \right>  &=& 0\label{eq:Gamma_uv1}  \\
    \left< \tilde{\mu}_M \middle|\Big(-i\hbar\sum_{I}({\bm X}_{I}-\bm G) \times {\hat{\bm {\Gamma}}}_{I} + \sum_I \hat{\Theta}_I (\hbr-\bm{G})\times {\hat {\bm k}_I} \Big)\middle| \tilde{\nu}_N \right> &=& 0\label{eq:Gamma_uv3}
\\
    \left< \tilde{\mu}_M \middle|\Big[-i\hbar\sum_{J}\pp{}{{X}_{J\beta}} + \hat{p}_\beta, {\hat {{\Gamma}}}_{I\alpha}\Big]\middle| \tilde{\nu}_N \right> &=& \frac{ie}{2\hbar} (\bB\times \mathbf{X}_{MN})_\beta { \tilde {{\Gamma}}}^{I\alpha}_{\mu\nu}\nonumber\\\label{eq:Gamma_uv2}\\
     \left< \tilde{\mu}_M \middle|\Big[-i\hbar\sum_{B}\left((\bm{X}_{B}-\bG) \times\pp{}{\bm{X}_B}\right)_{\gamma} + \hat{l}_{\gamma} + \hat{s}_{\gamma}, \hat{{\Gamma}}_{I \delta}\Big] \middle| \tilde{\nu}_N \right>
     &=& i\hbar \sum_{\alpha} \epsilon_{\alpha \gamma \delta} \tilde{{\Gamma}}^{I \alpha}_{\mu\nu}\label{eq:Gamma_uv4}
\end{eqnarray}
Above,  Eqs. \ref{eq:Gamma_uv1} and \ref{eq:Gamma_uv3} are phase conventions for the atomic orbitals under translations and rotations, and Eqs. \ref{eq:Gamma_uv2} and \ref{eq:Gamma_uv4} dictate that the $\Gamma$ must transform correctly under
translation and rotation to give translationally and rotationally invariant energies. 

Now, when comparing Eqs. $\mansia, \mansib, \mansic$, and $\mansid$ of Paper I with Eqs. \ref{eq:Gamma_uv1}-\ref{eq:Gamma_uv4} above,  an important difference arises in the third constraint due to our use of GIAOs. Namely, given that $\tilde {\bm \Gamma}_{\mu\nu}$ is defined in a GIAO basis, translation of the system leads to  the $\tilde {\bm \Gamma}_{\mu\nu}$ term picking up a phase under translation (just as for the matrix element of $\tilde{\bm{\pi}}$ in Eq. \ref{eq:piuv} above).  

\subsection{Proof that $E_{PS}$ is Independent of the Choice of $\bG$ }\label{ETF-ERF}

Before we can assert that $E_{PS}$ is invariant to $\bG$, it will be helpful to compute the matrix elements of $\hat{\bGamma}$ in a GIAO basis. To that end, recall that 
\begin{eqnarray}
\hat{\bGamma} = \hat{\bGamma}' + \hat{\bGamma}''
\end{eqnarray}

\subsubsection{GIAO Matrix Elements of $\bGamma'$}

Let us begin with $\tilde{\bGamma}'$ and evaluate Eq. \ref{eq:etf} above in a basis of GIAOs:
\begin{eqnarray}
\left< \tilde \mu_M \middle|\hat{\Gamma}'^{A\beta }\middle|\tilde \nu_N \right> &= &\frac{1}{2i\hbar}\left< \tilde \mu_M \middle|(\hat{\Theta}_I{{\hat{\pi}_\beta}}+{{\hat{\pi}_\beta}}\hat{\Theta}_I)\middle| \tilde \nu_N \right> + \frac{ie}{\hbar} \left< \tilde \mu_M \middle|(\bB\times (\hbr-\bX_I))_\beta\hat{\Theta}_I\middle| \tilde \nu_N \right>  \\&=& \frac{1}{2i\hbar}\left< \tilde \mu_M \middle|(\hat{\Theta}_I\hat{p}_{\beta}+ \hat{p}_{\beta}\hat{\Theta}_I)\middle| \tilde \nu_N \right>  - \frac{ie}{2\hbar}\left< \tilde \mu_M \middle|(\bB \times (\hbr-\bG))_\beta\hat{\Theta}_I\middle| \tilde \nu_N \right>\nonumber\\&&+ \frac{ie}{\hbar} \left< \tilde \mu_M \middle|(\bB\times (\hbr-\bX_I))_\beta\hat{\Theta}_I\middle| \tilde \nu_N \right> \\&=& -\frac{1}{2}\left< \mu_M \exp{(-\frac{ie}{\hbar}\mathbf{A}_M\cdot \br)} \middle|(\hat{\Theta}_I\nabla_{\beta}+ \nabla_{\beta}\hat{\Theta}_I)\middle| \exp{(-\frac{ie}{\hbar}\mathbf{A}_N\cdot \br)} \nu_N \right> \nonumber \\&&-\frac{ie}{2\hbar}\left< \mu_M \exp{(-\frac{ie}{\hbar}\mathbf{A}_M\cdot \br)} \middle|(\bB \times (\hbr-\bG))_\beta\hat{\Theta}_I\middle| \exp{(-\frac{ie}{\hbar}\mathbf{A}_N\cdot \br)} \nu_N \right> \nonumber\\&&+ \frac{ie}{\hbar} \left< \mu_M \exp{(-\frac{ie}{\hbar}\mathbf{A}_M\cdot \br)} \mu\middle|(\bB\times (\hbr-\bX_I))_\beta\hat{\Theta}_I\middle|\exp{(-\frac{ie}{\hbar}\mathbf{A}_N\cdot \br)} \nu_N \right>\nonumber\\
\end{eqnarray}
Thereafter, we can act the electronic momentum operator ($\nabla$) on the GIAOs to pull out the exponential factor so that we can express $\bm{\hat{\Gamma}}$ more easily in terms of the AO basis:
\begin{eqnarray}
\left< \tilde \mu_M \middle|\hat{\Gamma}'^{A\beta }\middle|\tilde \nu_N \right> &=& -\frac{1}{2}\left< \mu_M \middle|\exp{(\frac{ie}{\hbar}\mathbf{A}_{MN}\cdot \hbr)}\Big(\hat{\Theta}_I\nabla_{\beta}+ \nabla_{\beta}\hat{\Theta}_I\Big)\middle| \nu_N \right> \nonumber \\&&
+\frac{ie}{2\hbar}\left< \mu_M \middle|\exp{(\frac{ie}{\hbar}\mathbf{A}_{MN}\cdot \hbr)}\Big(\bB \times (\frac{\bX_M+\bX_N}{2}-\bG)\Big)_\beta\hat{\Theta}_I\middle|\nu_N \right>
\nonumber \\&&-\frac{ie}{2\hbar}\left< \mu_M \middle|\exp{(\frac{ie}{\hbar}\mathbf{A}_{MN}\cdot \hbr)}\Big(\bB \times (\hbr-\bG)\Big)_\beta\hat{\Theta}_I\middle|\nu_N \right> \nonumber \\&&+\frac{ie}{\hbar} \left< \mu_M \middle|\exp{(\frac{ie}{\hbar}\mathbf{A}_{MN}\cdot \hbr)}\Big(\bB\times (\hbr-\bX_I)\Big)_\beta\hat{\Theta}_I\middle| \nu_N \right>\\&=& -\frac{1}{2}\Big(\left< \mu_M \middle|\exp{(\frac{ie}{\hbar}\mathbf{A}_{MN}\cdot \hbr)}\hat{\Theta}_I\middle|\nabla_{\beta} \nu_N \right> -\left< \nabla_{\beta} \mu_M \middle|\exp{(\frac{ie}{\hbar}\mathbf{A}_{MN}\cdot \br)}\hat{\Theta}_I\middle| \nu_N \right> \Big)\nonumber\\&&-\frac{ie}{2\hbar}\left< \mu_M \middle|\exp{(\frac{ie}{\hbar}\mathbf{A}_{MN}\cdot \hbr)}\Big(\bB \times (\hbr-\frac{\bX_M+\bX_N}{2})\Big)_\beta\hat{\Theta}_I\middle|\nu_N \right> \nonumber \\&&+\frac{ie}{\hbar} \left< \mu_M \middle|\exp{(\frac{ie}{\hbar}\mathbf{A}_{MN}\cdot \hbr)}\Big(\bB\times (\hbr-\bX_I)\Big)_\beta\hat{\Theta}_I\middle| \nu_N \right>\label{eq:etf_giao}
\end{eqnarray}
The final expression for the dressed ETF term in Eq. \ref{eq:etf_giao} does not depend on the gauge $\bG$.  

\subsubsection{GIAO Matrix Elements of $\bGamma''$}
Next, let us turn to $\tilde \bGamma''$.
For short hand, notice that we can write Eq. \ref{eq:erf} as:
\begin{eqnarray}
   \tilde{\bm{\Gamma}}_I^{''} &=& \sum_{J}  \zeta_{IJ}\left(\bm{X}_I -\bm{X}^0_{J}\right)\times \left(\bm{K}_J^{-1}\tilde{\bm{J}}^J_{\mu\nu}\right)\label{eq:erf_giao_defn}
\end{eqnarray}
where
\begin{eqnarray}
    \tilde{\bm{J}}^I_{\mu\nu} &=& \frac{1}{2i\hbar}\left<\tilde \mu_M\middle|(\hbm{r}-\bm X_I)\times(\hat{\Theta}_I \bm{\hat k}_{I} + \bm{ \hat k}_{I}\hat{\Theta}_I)\middle|\tilde\nu_N\right>
\end{eqnarray}
Here,  $\tilde{\bm{J}}^I$ is effectively a pseudo-angular momentum. The matrix elements of $\tilde{\bm{J}}^I$   can be expressed in a GIAO basis as follows:
\begin{align}
    \tilde{J}^{I \alpha}_{\mu\nu} =& \frac{1}{2i\hbar} \left< \tilde \mu_M \middle|\sum_{I\beta\gamma}\epsilon_{\alpha\beta\gamma} (\hbr- \bX_I)_\beta (\hat{\Theta}_I {\hat k^{I}_\gamma} + {\hat k^{I}_\gamma}\hat{\Theta}_I) \middle|\tilde \nu_N \right>\\ = &-\sum_{I\beta\gamma}\epsilon_{\alpha\beta\gamma} \Bigg(\frac{1}{2}\Big[\left< \mu_M \middle|\exp{(\frac{ie}{\hbar}\mathbf{A}_{MN}\cdot \br)}(\hbr- \bX_I)_\beta \hat{\Theta}_I\middle| \nabla_{\gamma}\nu_N \right> \nonumber \\
    &-\left< \nabla_{\gamma} \mu_M\middle|\exp{(\frac{ie}{\hbar}\mathbf{A}_{MN}\cdot \br)} (\hbr- \bX_I)_\beta \hat{\Theta}_I\middle| \nu_N \right> \Big]\nonumber\\&-\frac{ie}{2\hbar}\left< \mu_M \middle|\exp{(\frac{ie}{\hbar}\mathbf{A}_{MN}\cdot \br)}(\hbr- \bX_I)_\beta(\bB \times (\hbr-\frac{\bX_M+\bX_N}{2}))_\gamma\hat{\Theta}_I\middle|\nu_N \right> \nonumber\\
    &+\frac{ie}{\hbar} \left< \mu_M \middle|\exp{(\frac{ie}{\hbar}\mathbf{A}_{MN}\cdot \br)}(\hbr- \bX_I)_\beta(\bB\times (\hbr-\bX_I))_\gamma\hat{\Theta}_I\middle| \nu_N \right>\Bigg)\label{eq:Juv}
\end{align}
As expected, this expression (and the form of $\bGamma''$)  does not depend on the gauge origin $\bG$. Therefore,  all together, the total phase-space energy does not depend on $\bG$, i.e.,
\begin{eqnarray}
    \left(\frac{\partial E^{PS}}{\partial \bG}\right)_{\bm{\Pi}^{\it{eff}}} = 0 
    \label{eq:depsdg}
\end{eqnarray}

\subsection{Proof of translational invariance for $E_{PS}$}\label{gamma_trans}

We now have the tools to prove that $E_{PS}$ is translationally invariant.
According to their definitions and matrix elements in Eqs. \ref{eq:etf_giao}-\ref{eq:Juv} above, it is straightforward  to show that, in a GIAO basis,  $\tilde {\bm \Gamma}_{\mu\nu}'$ and $\tilde {\bm\Gamma}_{\mu\nu}''$ transform under translation (and pick up a phase factor [in a manner analogous to Eq. \ref{translate_hgiao} above]) and satisfy Eq. \ref{eq:Gamma_uv2}.   The proof is obvious for $\tilde {\bm\Gamma}_{\mu\nu}'$; for the case of $\tilde {\bm\Gamma}_{\mu\nu}''$, note that the phase factor arises from transforming $\tilde{J}^{I \alpha}_{\mu\nu}$, and recognizing that 
 the terms $\bm K_{J}$ and $\left(\bm{X}_I-\bm{X}_{J}^0\right)$ 
 (from Eq. \ref{eq:erf_giao_defn})
are invariant to translation \cite{tian_erf,tao2024basis}. Thus, $E_{PS}$ must be translationally invariant in the same way as $E_{BO}$ (see Sec. \ref{giao_energy_trans}).

\subsection{Proof of rotational invariance for $E_{PS}$ around G}\label{gamma_rot}

Next we turn to rotations. Given the argument for the rotational invariance of $E_{BO}$ in Sec. \ref{giao_energy_rot} above, and given the new terms in the Hamiltonian in Eq. \ref{eq:PSH}, one can prove that 
$E_{PS}$ is rotationally invariant simply by demonstrating that the pseudomomentum vectors transform correctly under rotation, i.e.,
\begin{eqnarray}
    \label{eq:kI_rot}
    \hat{\bm{k}}^I_{\bar{\mu}\bar{\nu}}(\bm{R} \bm X) &=& \bm{R} \hat{\bm{k}}^I_{\mu\nu}(\bm X) \\
    (\hbr\times \hat{\bm{k}}^I)_{\bar{\mu}\bar{\nu}}(\bm{R} \bm X) &=& \bm{R} (\hbr\times \hat{\bm{k}}^I)_{\mu\nu}(\bm X)\label{eq:kIL_rot}
\end{eqnarray}
Eqs. \ref{eq:kI_rot}-\ref{eq:kIL_rot} are proven in Appendix \ref{rot_proof}, and reflect the fact that
rotating the physical system is equivalent to 
rotating both the pseudomomentum vector $\hat{\bm{k}}$ as well as the basis function around their atomic centers. Note that $\left(\bm{X}_I-\bm{X}_{J}^0\right)$  and $\bm K_{J}$ also transform like vectors (with the proper rotational properties\cite{tian_erf,tao2024basis}).
Thus, from Eqs. \ref{eq:kI_rot}-\ref{eq:kIL_rot}, it follows that: 
\begin{eqnarray}
    \tilde{\bm{\Gamma}}^{I'}_{\bar{\mu}\bar{\nu}}(\bm{R} \bm X) &=& \bm{R}  \tilde{\bm{\Gamma}}^{I'}_{{\mu}{\nu}}(\bm X) \\
    \tilde{\bm{\Gamma}}^{I''}_{\bar{\mu}\bar{\nu}}(\bm{R} \bm X) &=& \bm{R} \tilde{\bm{\Gamma}}^{I''}_{{\mu}{\nu}}(\bm X)
\end{eqnarray}
which implies that $E_{PS}$ is rotationally invariant.

\section{Conservation of Pseudomomentum and Angular Momentum}\label{pm-am-conserve}

In Paper I, we have shown that Hamiltonian dynamics along an eigenstate of $\hat{H}_{PS}$ conserves pseudomomentum and angular momentum. Now, it is crucial to emphasize that, in establishing such conservation laws, in Eqs. $\mansie$ and $\mansif$ of Paper 1, we required only translational and rotational invariance of $E_{PS}$. At no point did we explicitly require diagonalization of $\hat{H}_{PS}$ or explicit gauge invariance. Instead, by contrast, we merely noticed that gauge invariance and diagonalization led to translational and rotational invariance, which in turn led to  pseudomomentum and angular momentum conservation.

Now, the idea of GIAOs is to use the observation above.  Namely, even without exact diagonalization and/or explicit gauge invariance, one can use GIAOs to recover translational and rotational invariance, and this invariance will further lead to momentum conservation. 
 To that end, and for the sake of completeness, we will now delineate the relevant expressions of the pseudomomentum and angular momentum, and outline why they are conserved in phase-space theory even in a finite AO basis (where we lose  the  gauge invariance from Eq. \ref{eq:depsdg} because we do not directly diagonalize the full $\hat{H}_{PS}$). 
Explicit derivations (following Paper I line-by-line) are presented in the Appendix \ref{am-pm_conserve} for the reader who prefers AO-based dynamics, but the formalism is identical to what we have presented in Paper I. 



\subsection{Equations of Motion}
In order to prove all flavors of momentum conservation, it will be necessary to derive the appropriate equations of motion. Using Eqs. \ref{eq:pssh_e}-\ref{eq:std_e} above, we can find phase-space equations of motion through Hamilton's principle:
\begin{eqnarray}
    \dot{X}_{I\alpha}&=&\left(\frac{\partial E_{PS}}{\partial P_{I\alpha}}\right)_{\bm X}\\
    &=& \frac{ \Pi^{\textit{eff}}_{I\alpha}}{M_{I}} - i\hbar\sum_{\mu\nu}\frac{\tilde{D}_{\nu\mu}{\tilde{\Gamma}}^{I\alpha}_{\mu\nu}}{M_{I}}\label{eq:xdot}\\
    \dot{P}_{I\alpha}&=&- \left(\frac{\partial E_{PS}}{\partial X_{I\alpha}}\right)_{\bm P}  \\
        &=&\sum_{J\beta} \frac{\Pi_{J\beta}^{\textit{eff}}}{M_J}\frac{\partial }{\partial X_{I\alpha}}\Bigg(- \frac{1}{2} q^{\textit{eff}}_J(\bX)(\bB\times (\bX_J-\bG))_{\beta}\Bigg)_{\bm P}  
        -\left(\frac{\partial V_{PS}}{\partial X_{I \alpha}}\right)_{\bm P} 
        \label{eq:pdot}
\end{eqnarray}
We can further decompose the gradient of the phase-space potential energy (with  fixed $\bP$)  into two terms:
one involving the gradient of the vector potential, and the other involving the gradient of the phase-space potential energy (with $\bm{\Pi}^{\textit{eff}}$):
\begin{eqnarray}
 \left(\frac{\partial V_{PS}}{\partial X_{I \alpha}}\right)_{\bm P} &=&  - i\hbar \sum_{\nu\mu J\beta} \frac{\tilde{D}_{\nu\mu}}{M_J}\frac{\partial}{\partial X_{I\alpha}}\Big(- \frac{1}{2} q^{\textit{eff}}_J(\bX) (\bB\times (\bX_J-\bG))_{\beta} \Big)_{\bm P}  {\tilde{\Gamma}}^{J\beta}_{\mu\nu}-
\left(\frac{\partial V_{PS}}{\partial X_{I \alpha}}\right)_{{\bm\Pi}^{\textit{eff}}}\nonumber\\\label{eq:Estdgrad}
\\
     \left(\frac{\partial V_{PS}}{\partial X_{I \alpha}}\right)_{{\bm\Pi}^{\textit{eff}}} & = & 
     \sum_{\nu\mu}\tilde{D}_{\nu\mu}\Bigg(\frac{\partial \tilde{h}_{\mu\nu}}{\partial X_{I\alpha}} -\sum_{J\beta}\frac{i\hbar \Pi^{\textit{eff}}_{J\beta}}{M^{J}} \frac{\partial {\tilde{\Gamma}}^{J\beta}_{\mu\nu}}{\partial X_{I\alpha}} -\hbar^2 \sum_{J\beta}\frac{1}{2M_J}\frac{\partial ({\tilde{\Gamma}}^2_{J\beta})_{\mu\nu}}{\partial X_{I\alpha}}\Bigg)  \nonumber \\&&+ \sum_{\mu\nu\lambda\sigma}\tilde{d}_{\nu\mu\lambda\sigma}\frac{\partial \tilde{g}_{\nu\mu\sigma\lambda} }{\partial X_{I\alpha}} - \sum_{\mu\nu} \tilde W_{\nu\mu}\frac{\partial \tilde {S}_{\mu\nu}}{\partial X_{I\alpha}}   \label{eq:gradEpi}
\end{eqnarray}
The last term (proportional to $\tilde W_{\mu\nu}$) follows by differentiating the one and two electron density matrices\cite{szabo1996modern}; as is well known, one can use Wigner's rule to avoid calculating orbital response explicitly (but a new term does still arise because of the dependence of the overlap matrix $\tilde S_{\mu\nu}$ on nuclear position).

\subsection{Pseudomomentum Conservation}\label{pmconservedefn}
The proof for pseudomomentum conservation now follows similarly as in Paper I. We posit that the conserved pseudomomentum is the sum of a nuclear and electronic observable:
\begin{eqnarray}
  K^\alpha_{mol} &=&
  K^\alpha_{n} +
  k^{\alpha}_{e}
  \\
  K^\alpha_{n} &=&
  \sum_I {P_{I\alpha}}+ \frac{1}{2}\sum_I q^{\textit{eff}}_I(\bX)(\bB\times (\bX_I-\bG))_{\alpha}-i\hbar \sum_{I\mu\nu} \tilde{D}_{\nu\mu}\tilde{{\Gamma}}_{\mu\nu}^{I\alpha}  \\ 
  k^{\alpha}_{e} &=& \sum_{I\mu \nu} \hat{\Theta}_I\tilde{D}_{\nu \mu}  \tilde{k}^{I\alpha}_{\mu\nu}
\label{eq:pseudomom0}
\end{eqnarray}
As in Paper I, one can prove that$\frac{dK^\alpha_{mol}}{dt} = 0$ by realizing that  $\sum_I \left(\frac{\partial V_{PS}}{\partial X_{I \alpha}}\right)_{\bm\Pi} = 0$ (which we have already shown in Sec. \ref{gamma_trans} using GIAOs). 
Note that, while the pseudomomentum is conserved, the conserved quantity does depend on $\bG$, unlike the case of energy.

\subsection{Angular Momentum Conservation}\label{amconservedefn}
Similarly, the total canonical angular momentum (which is sum of canonical angular momentum and the canonical pseudomomentum like term) 
follows the definition in Paper I:
\begin{eqnarray}
    L_{mol}^z &=& L_n^z + L_e^z\\
    L_n^z &=& \sum_I \Bigg((\bX_I-\bG) \times \Big( \bm{\Pi}^{\textit{eff}}_{I} - i\hbar \sum_{\mu\nu}\tilde{D}_{\nu\mu}\bm{\tilde{\Gamma}}^{I}_{\mu\nu}+\frac{q_I^{\textit{eff}}(\bX)}{2}(\bB\times (\bX_I-\bG))\Big)\Bigg)_z\\
    L_e^z &=&  \sum_{I\mu\nu}  \hat{\Theta}_I\tilde D_{\nu\mu}{(( \hbr- \bG)\times \tilde{\bm{k}}^I_{\mu\nu})}_{z} \\    
    L_{mol}^z &=& \sum_I \Bigg((\bX_I-\bG) \times \Big( \bm{\Pi}^{\textit{eff}}_{I} - i\hbar \sum_{\mu\nu}\tilde{D}_{\nu\mu}\bm{\tilde{\Gamma}}^{I}_{\mu\nu}+\frac{q_I^{\textit{eff}}(\bX)}{2}(\bB\times (\bX_I-\bG))\Big)\Bigg)_z \nonumber\\&&+ \sum_{I\mu\nu}  \hat{\Theta}_I\tilde D_{\nu\mu}{(( \hbr- \bG)\times \tilde{\bm{k}}^I_{\mu\nu})}_{z}  
\end{eqnarray}
Furthermore, as in Paper I, one can  prove that this 
quantity is conserved by noticing that $E_{PS}$ is rotationally invariant, i.e.,
\begin{eqnarray}
\frac{d}{dt} L^{z}_{mol}
&=&  -\sum_{I\beta\gamma}\epsilon_{z\beta\gamma}\Big(P_{I\beta} \left(\frac{\partial E_{PS}}{\partial P_{I\gamma}}\right)_{\bm X}+ (X_{I\beta}-G_\beta) \left(\frac{\partial E_{PS}}{\partial X_{I\gamma}}\right)_{\bm P}\Big)
\end{eqnarray}
See Sec. \ref{gamma_rot} and Appendix \ref{rot_proof}. 

\section{Preliminary Results: Momentum Dependence of the Potential Energy Surfaces}\label{results}

\begin{figure} [ht]
\includegraphics[scale = 0.3]{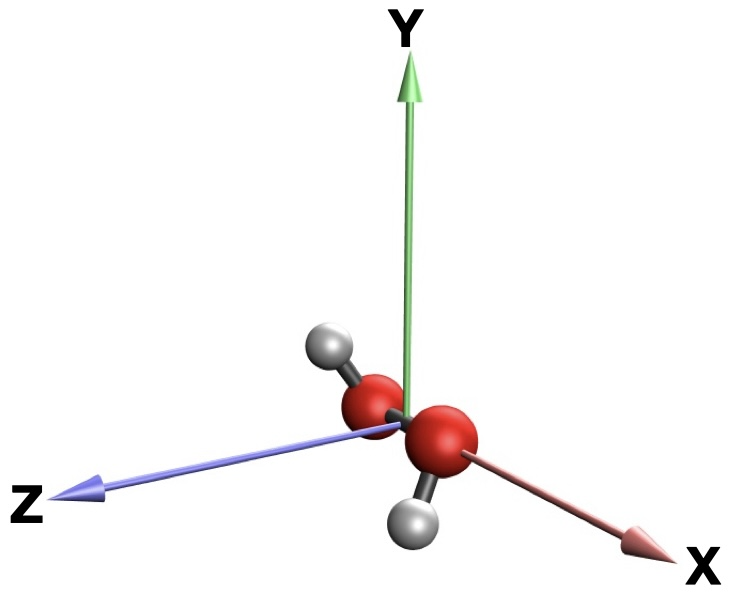}
\caption{Orientation of $H_2O_2$ relative to the $x,y,z$ axes.}
\label{fig:h202geom}
\end{figure}

The theory above has been implemented in a developmental version of the Q-Chem electronic structure package\cite{qchem}. For the calculations below, we ignore the 
$\bGamma^2$ term in Eq. \ref{eq:std_e} (which is small).
Our test example is a chiral molecule, $H_2O_2$, shown in Fig. \ref{fig:h202geom}; note the direction of the $x,y$, and $z$ axes. 

\begin{figure} [ht]
\centering
\begin{subfigure}[b]{0.495\linewidth}
\centering
\includegraphics[width=1.0\textwidth]{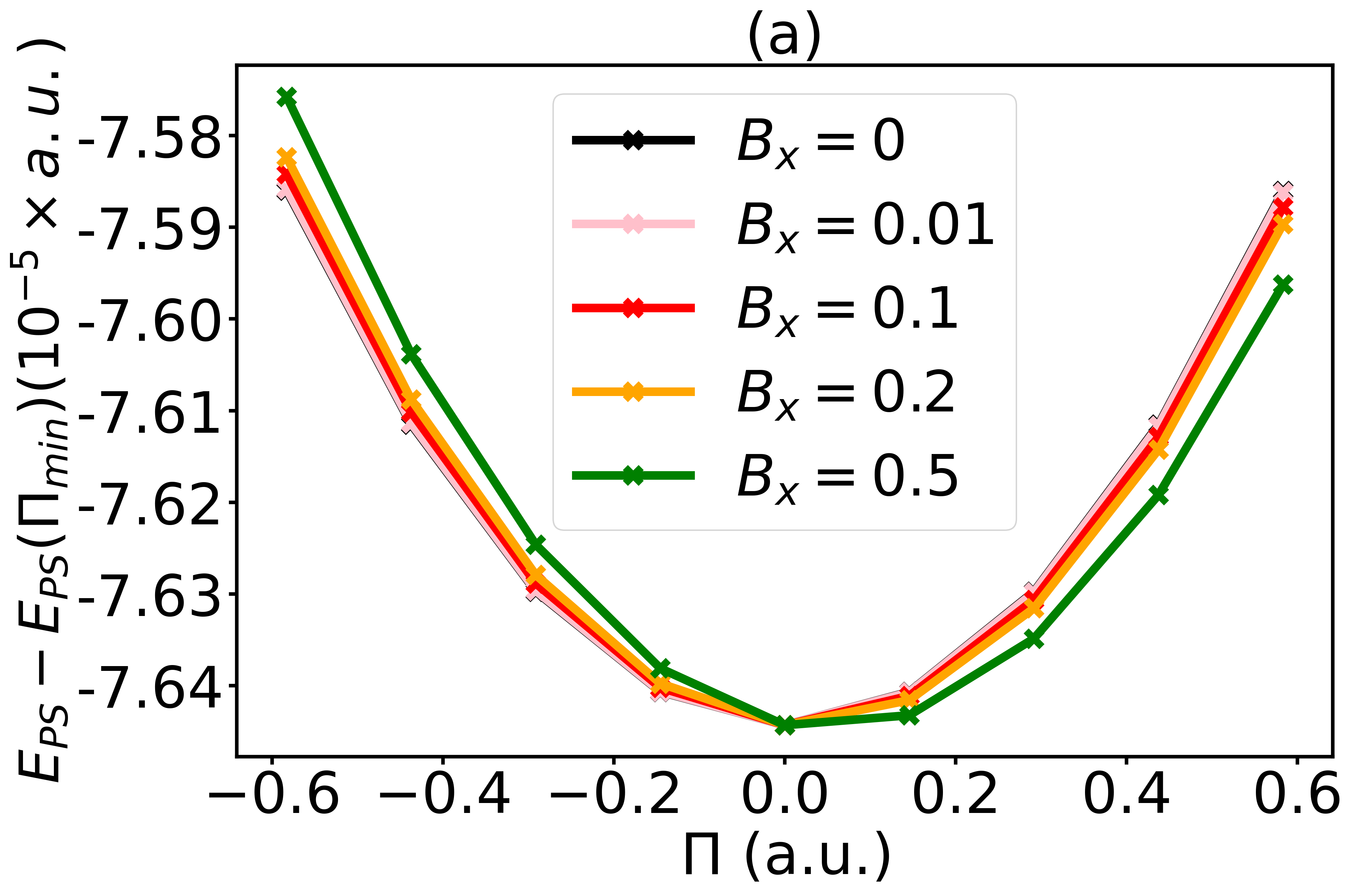}
\end{subfigure}
\begin{subfigure}[b]{0.495\linewidth}
\centering
\includegraphics[width=1.0\textwidth]{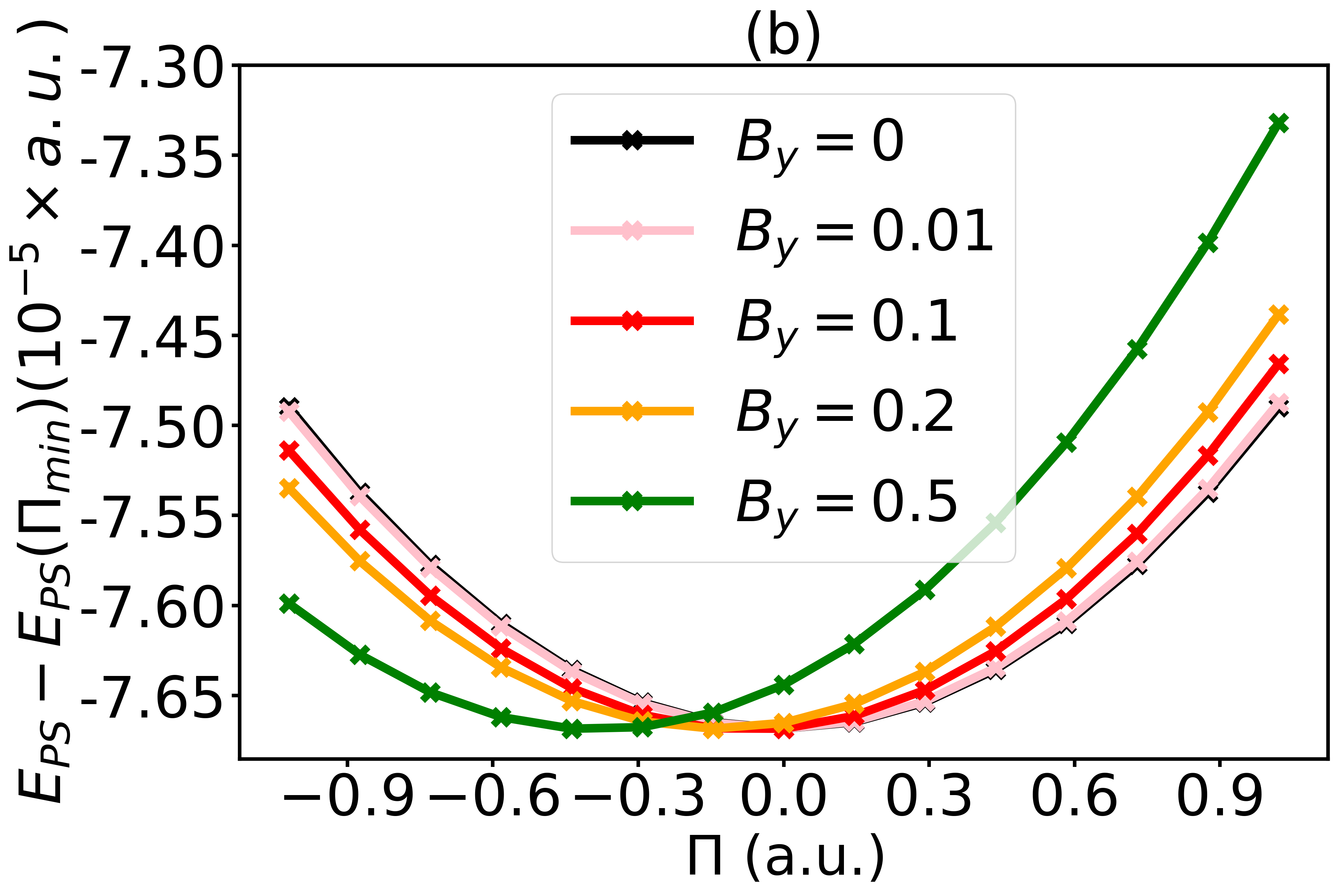}
\end{subfigure}
\begin{subfigure}[b]{0.495\linewidth}
\centering
\includegraphics[width=1.0\textwidth]{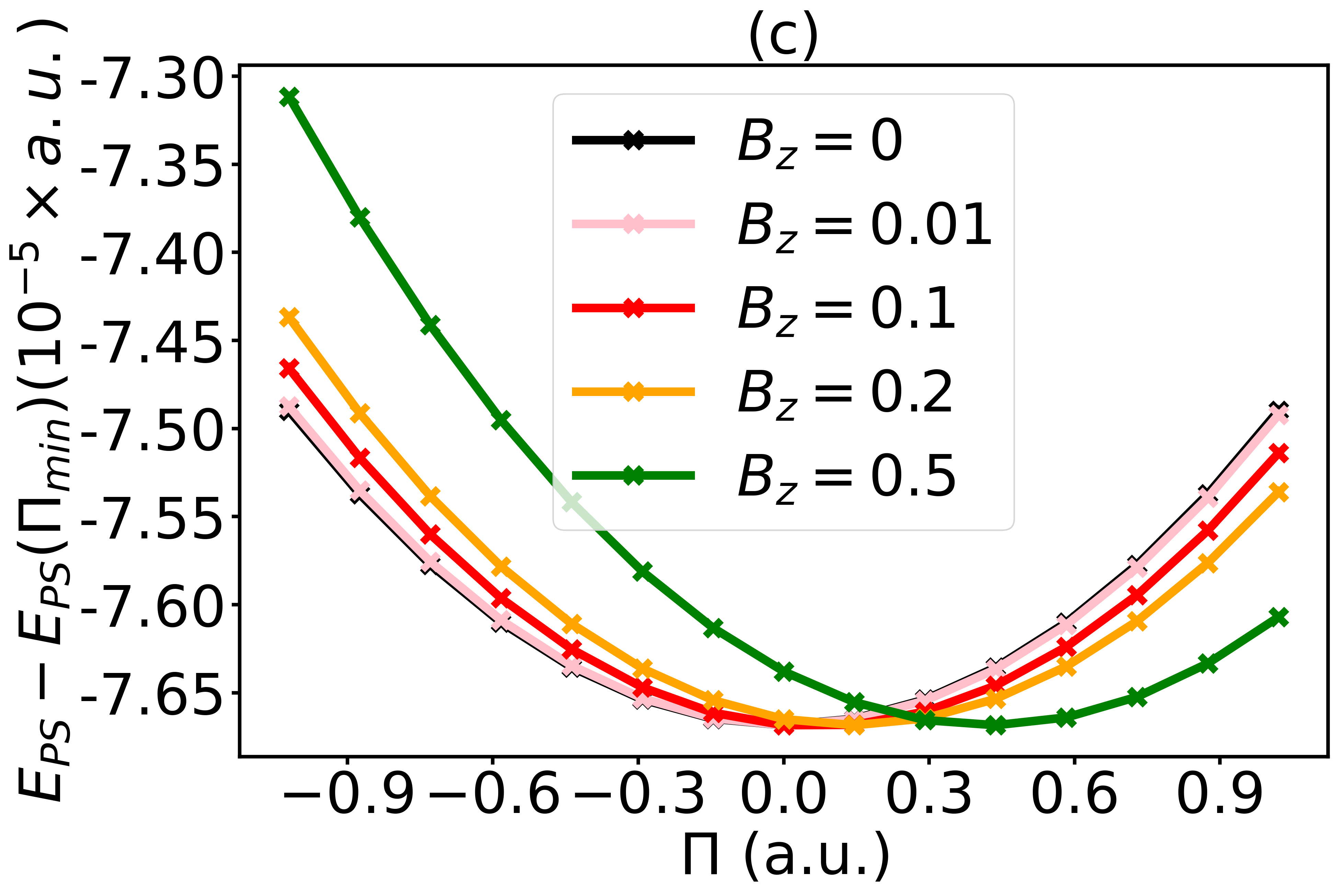}
\end{subfigure}
\begin{subfigure}[b]{0.495\linewidth}
\centering
\includegraphics[width=1.0\textwidth]{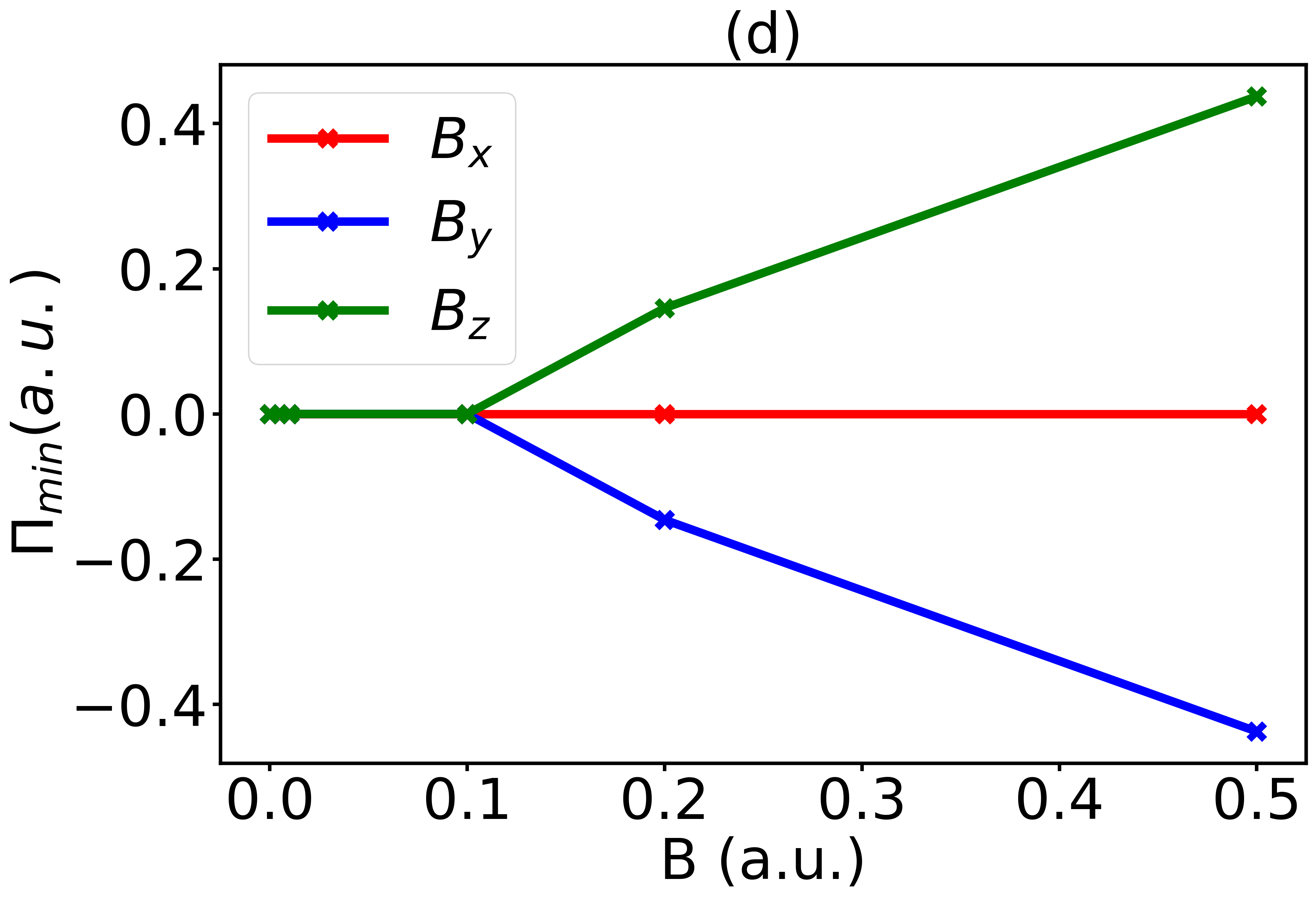}
\end{subfigure}
\begin{subfigure}[b]{0.495\linewidth}
\centering
\includegraphics[width=1.0\textwidth]{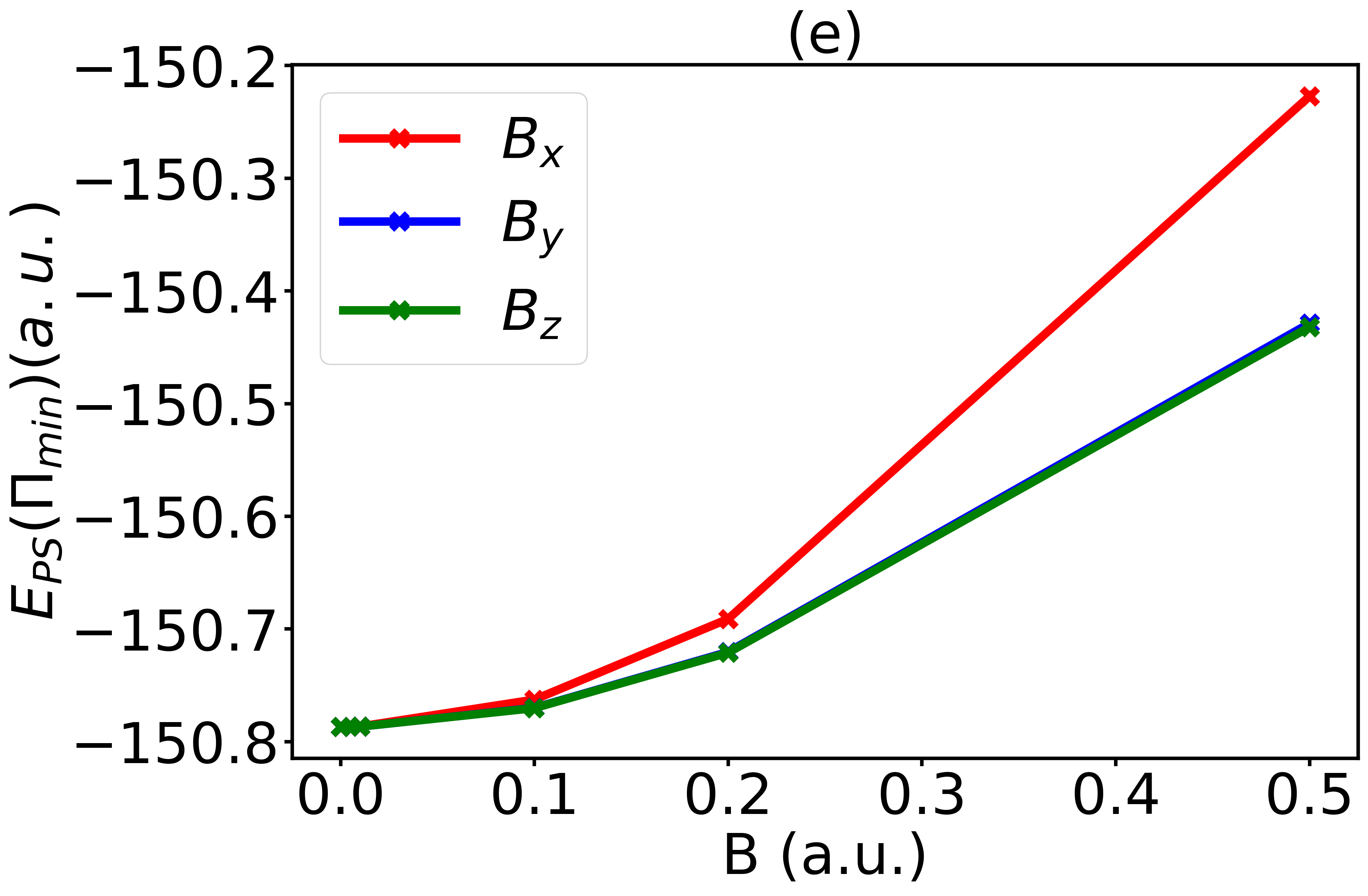}
\end{subfigure}
\caption{The phase space energy as a function of the nuclear kinetic momentum ($\bm{\Pi}^{\textit{eff}}$) for varying magnetic fields applied in the $x$(a), $y$(b) and $z$(c) directions. Each curve has been shifted vertically so that all of the curves (with different magnetic fields) have the same minimum energy. (d) The nuclear kinetic momentum (at the minimum phase space energy) as a function of the magnetic field; note that for $\bB \ne 0$, the minimum energy occurs for $\bm{\Pi}^{\textit{eff}}_{min} \ne 0$. (e) The minimum phase-space energy as a function of magnetic field.  }
\label{fig:res}
\end{figure}

Within BO theory, the energy of a given molecular system is separable between the nuclear position and nuclear momentum; the nuclear kinetic energy $\bm{\Pi}^2/2{\bm M}$ is simply added on top of the potential energy. Within phase-space theory, however, there is no such separation. To that end, 
in Fig. 
\ref{fig:res}(a),(b) and (c), we plot the relative potential energy surfaces as a function of $\bm{\Pi}^{\textit{eff}}$ for different magnetic field strengths.  (Recall that 1 a.u. of magnetic field is $2.35 \times 10^5$ Tesla, so these are large fields.)   From this figure, the most interesting feature that emerges is that, while the dependence of the energy on $\bm{\Pi}^{\textit{eff}}$ remains quadratic, for $\bB_y \ne 0$ or $\bB_z \ne 0$, we find that $\bm{\Pi}_{min}^{\textit{eff}} \ne 0$.  In Fig. \ref{fig:res}(d), we plot how the minimum in $\bm{\Pi}^{\textit{eff}}$ change as a function of the $\bB$ field; as the magnetic field increases in magnitude, the minimum moves farther and farther away from zero. Lastly, in Fig. \ref{fig:res}(e), we plot the minimum value of  $E_{PS}$ for each $\bB$ field. Because $H_2O_2$ is diagmagnetic, note that the ground-state energy increases with magnetic field strength.  (This diamagnetic (vertical) shift in the minimum energy has been used to align the energy diagrams in Figs. (a)-(c) only so that we can more easily compare the horizontal shift in $\bm{\Pi}^{\textit{eff}}$ that is uniquely predicted by a phase-space electronic Hamiltonian.) Note that $1/\sigma^2$ defined in Eq. \ref{eq:thetadef} is used to parametrize the space around each atom based on fitted electronegativity has values 2.42 for $O$ atom and 3.79 for $H$ atom in $H_2O_2$ molecule.

Now, to interpret the finding that $\bm{\Pi}_{min}^{\textit{eff}} \ne 0$, it is crucial to emphasize that, at the minimum we do find that $\bm{\Pi}_{min}^{\textit{eff}}- \left<\Psi\middle|\hat{\bm{\Gamma}}\middle|\Psi\right> = 0$. In other words, according to phase-space theory, at the minimum energy point, the true kinetic momentum and the kinetic energy of the nuclei is zero -- the nuclei are not physically moving. 
That being said, however, the {\em electrons} do move as a result of the fact that $\bm{\Pi}_{min}^{\textit{eff}} \ne 0$ . In particular,  we know from Eqs. \ref{eq:etf} and \ref{eq:pseudolike} that: 
\begin{eqnarray}
    \frac{1}{i\hbar}\left< \Psi \middle| \hat{\bm{\pi}}_e\middle| \Psi \right> &=& \sum_I \left< \Psi \middle|\bm{\hat{\Gamma}}_I \middle| \Psi\right>  + \frac{1}{i\hbar}\sum_I\left< \Psi \middle| \hat{\Theta}_I  e\bB  \times (\hbr - \bX_I) \middle|\Psi \right>
\end{eqnarray}
and 
\begin{eqnarray}
    \frac{1}{i\hbar}\left< \Psi \middle| (\hbr\times \hat{\bm{\pi}}_e)\middle| \Psi \right> &=& \sum_I \left< \Psi \middle|(\hbr \times \bm{\hat{\Gamma}}_I) \middle| \Psi\right>  + \frac{1}{i\hbar}\sum_I\left< \Psi \middle| \hat{\Theta}_I  e\Big( \hbr\times (\bB  \times (\hbr - \bX_I))\Big) \middle|\Psi \right>\nonumber \\
\end{eqnarray}
Moreover, we expect $\left<\hat{\bGamma}\right>$ to grow with increasing $\bm{\Pi}^{\textit{eff}}$.
 \pagebreak
 
\begin{figure} [ht]
\centering
\begin{subfigure}[b]{0.495\linewidth}
\centering
\includegraphics[width=1.0\textwidth]{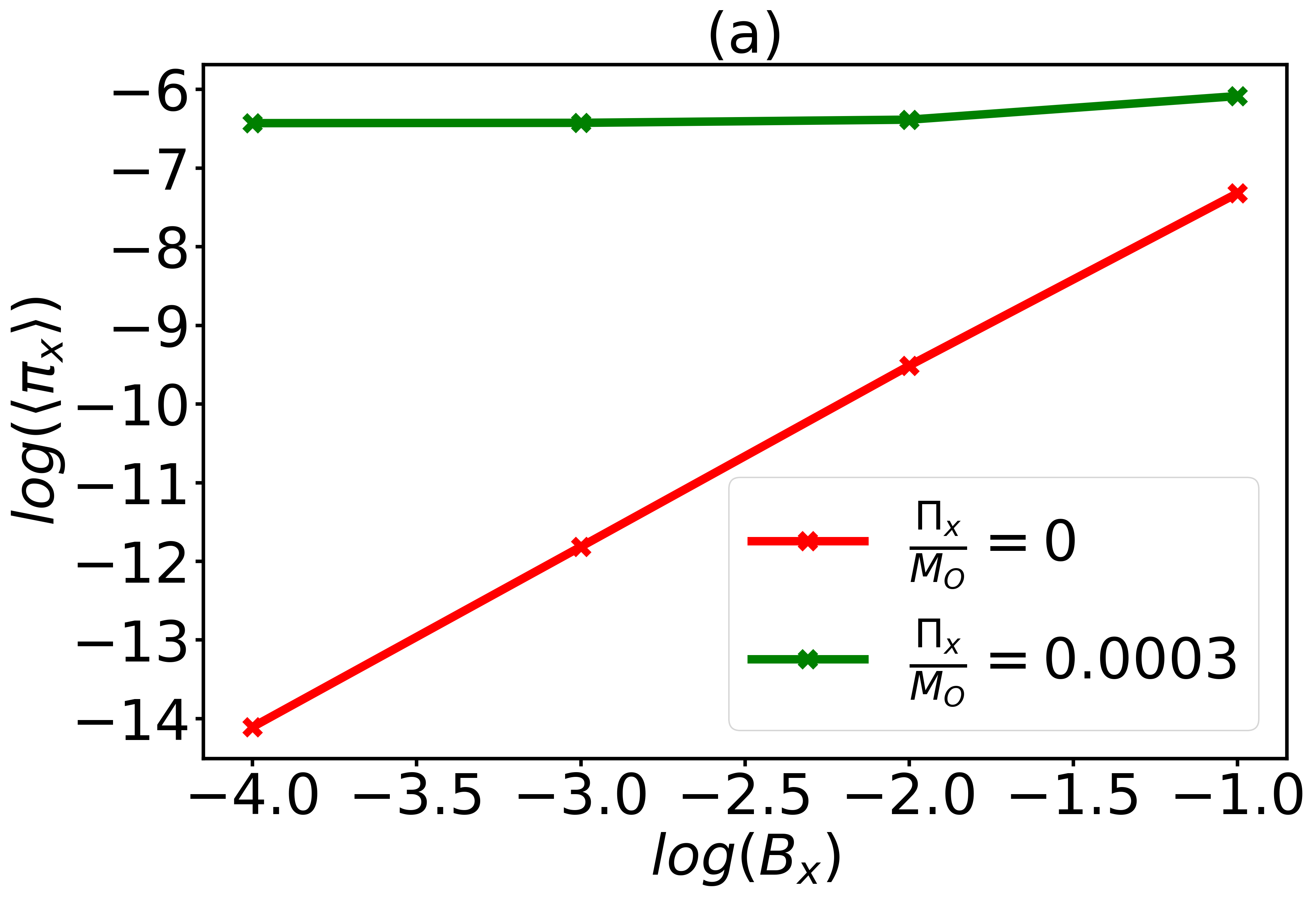}
\end{subfigure}
\begin{subfigure}[b]{0.495\linewidth}
\centering
\includegraphics[width=1.0\textwidth]{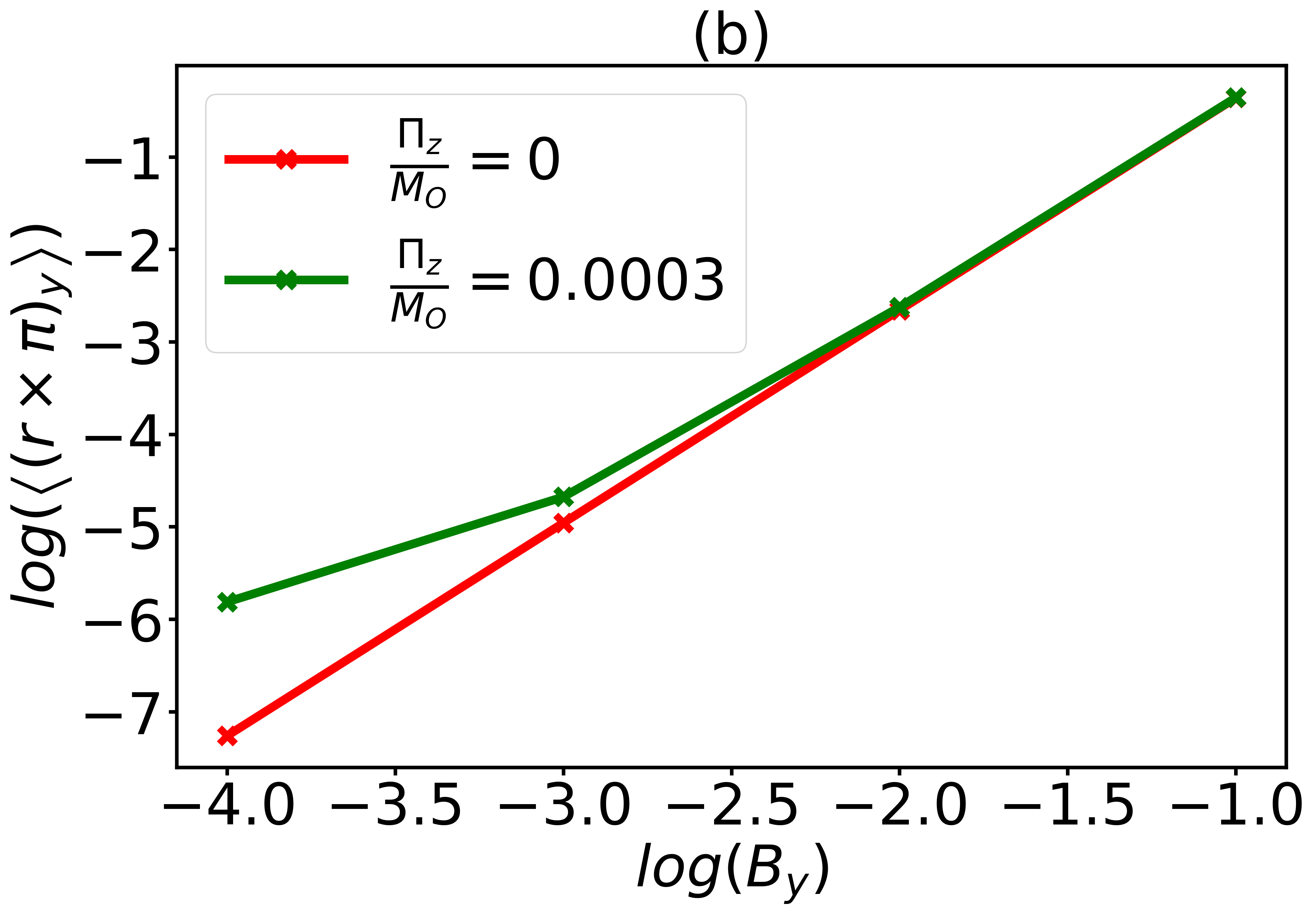}
\end{subfigure}
\caption{A plot  (a) $\langle \bm{\pi}_e \rangle$ and (b) $\langle \br\times \bm{\pi}_e \rangle$ as a function of magnetic field strength for two different nuclear kinetic momentum $\bm{\Pi}^{eff}$ using phase space theory with a a cc-pvdz basis and GIAOs (and ignoring the $\bGamma^2$ term in Eq. \ref{eq:std_e}). Note that, when $\bm{\Pi}^{eff} = 0$, we should find that that $\langle \bm{\pi}_e \rangle = 0$; any deviation from this result (in panel (a)) arises from an incomplete basis (which cannot be saved by using GIAOs). That being said, note the difference in y-axis values; these errors are small.  Turning to angular momentum (in panel (b)), there is no reason that  $\langle \br\times \bm{\pi}_e \rangle$ should vanish, and the curve in (b) shows that  indeed for small $\bB$-fields, the electronic angular momentum is dominated by the nuclear momentum. However, for large magnetic fields, the electronic angular momentum is dominated by the $\bB$-field itself. These results highlight how magnetic fields and nuclear motion can collectively combine (in interesting ways) to dictate electronic dynamics.}
\label{fig:res2}
\end{figure}

To partially quantify the resulting electronic dynamics,  in Fig. \ref{fig:res2}(a) and (b), we plot the linear ($\left< \hat{\bm{\pi}}\right>$) and angular ($\left< \hat{\bm r}\times \hat{\bm{\pi}}\right>$) electronic momentum.  Let us consider first the linear momentum. 
For ${\bm{\Pi}}^{\textit{eff}} \ne 0$, the linear momentum is roughly constant, independent of the size of the magnetic field. However, for a stationary hydrogen peroxide molecule ($\bm{\Pi}^{\textit{eff}} = 0$), the linear electronic momentum does grow with magnetic field. Unfortunately, the latter growth is an artifact of not using a complete basis (here we use cc-pvDZ with GIAOs) to correctly treat the magnetic field. After all, given that we ignore the $\bGamma^2$ term in Eq. \ref{eq:std_e} and that $\bm{\Pi}^{\textit{eff}} = 0$,  the phase space data in Fig. \ref{fig:res2}(a) should match with the BO data. Moreover, noting that 
\begin{eqnarray}
     -i\hbar\left< \Psi \middle|\frac{\hat{\bm{\pi}}_e}{m_e}\middle|\Psi \right> &=& \left< \Psi \middle|[\hat{H}_{BO},\hbr]\middle|\Psi \right> \\
     &=& \left< \Psi \middle|(E_{BO}\hbr-\hbr E_{BO})\middle|\Psi \right> \\
   &=& 0
     \label{eq:commmom}
\end{eqnarray}
it follows that, within BO theory, an electronic stationary state should always have zero electronic momentum. Thus, the increase of the red line in Fig. \ref{fig:res2}(a) must be basis set error. That being said, notice the small $y$-axis values: the error is not very large for small $\bB$ fields. 
 
Notwithstanding this caveat, the argument in Eq. \ref{eq:commmom}  holds only for linear (but not angular momentum); angular momentum cannot be written as the commutator of any observable with the Hamiltonian and should not vanish for stationary momentum provided $\bB \ne 0$. To that end, in Fig. \ref{fig:res2}(b), we plot the kinetic angular momentum as a function of magnetic field. Here, we see roughly what we expect to find (again, within the limits of an incomplete basis):  the expectation value of the relative angular momentum is dictated by the velocity for small magnetic fields but for very larger magnetic fields, the expectation value becomes dominated by the $\bB-$field strength.
Altogether, though we will require large basis sets for convergence, investigating exactly how electronic angular momentum is built from a combination of nuclear momentum and an external magnetic field  will be crucial as far as understanding magnetic field effects in the future. 

\section{Discussion and Conclusions}\label{conclude}

In the text above, we have specified how to calculate all of the relevant terms in 
Eq. \ref{eq:PSH} so that the reader can construct (in practice) a phase-space electronic Hamiltonian $\hat{H}(\bX,\bm{\Pi})$ for molecules in a magnetic field, parameterized by both nuclear position and momentum. As shown above, if we diagonalize this Hamiltonian to generate eigenstates  and eigenenergies, dynamics along the resulting eigenstates are guaranteed to conserve total pseudomomentum and angular momentum in the direction of the field. 
Moreover, if we minimize the energy of the system as a function of $\bm{\Pi}^{\textit{eff}}$, we can learn new physics insofar as $\bm{\Pi}^{\textit{eff}}_{min} \ne 0$. After all, as shown in Fig. \ref{fig:res2}(b), the electronic angular momentum is dictated by both the nuclear velocity of the atoms in the molecule as well as the strength of the external magnetic field.

Now, at the moment, our phase-space code is far from optimized, existing only in preliminary Q-Chem implementation.\cite{qchem}
Our next step is clearly to develop an efficient code, which should be possible given that phase-space codes require the inclusion of only one-electron terms on top of a BO Hamiltonian. Thereafter, we would like to compute efficient gradients and derivative couplings, and then begin to optimize geometries and run dynamics so to explore both structure as well as adiabatic and nonadiabatic effects as observed in the presence of magnetic fields. Magnetic field effects are well known in literature\cite{steiner1989} but notoriously difficult to model because they presumably depend on dynamical (rather than thermodynamical) mechanisms. In developing the present practical approach,  our motivation is to begin the process of understanding such magnetic field effects. To that end, one of our immediate aims will be to apply the current approach to study MCD and MVCD spectroscopy.\cite{fedotov2022magnetic,barron2009molecular,ganyushin2008first, sun2020relativistic}

Finally, given the opportunity to study magnetic field effects that correctly treat momentum, there are two obvious experiments that we would like to study in the near future. First,  there is currently a great deal of interest in understanding the chiral induced spin selectivity (CISS) effect\cite{doi:10.1021/acs.accounts.0c00485,doi:10.1021/jz300793y,Evers2022}, whereby electrons traveling through chiral media can be shown to have a strong preference -- even at room temperature\cite{kim2021chiral,alwan2023temperature}. Because spin is a form of angular momentum, we have a great deal of hope that studying electron transport with the current phase-space approach (that correctly conserves angular momentum) may reveal a new mechanism for the CISS effect\cite{scienceperspective}. Second, the essential physics of the Einstein-de Haas  effect\cite{edhass1,edhass2}  remains a key target for the current approach. For this application, the basic experiment is that, upon changing the magnetization of a ferromagnetic material, one finds that the material rotates as is necessary to conserve angular momentum. Obviously, if we have found a means to efficiently analyze and model how angular momentum is partitioned and exchanged between different forms of matter, the present approach has the potential to predict Einstein-de Haas physics at a molecular length scale\cite{ganzhorn2016quantum,wells2019microscopic,lemesheko:2015:prl}, demonstrating a deep connection between standard nonadiabatic approaches and magnetic field chemistry.

\section{Acknowledgments}
We thank Tanner Culpitt for helpful discussions about GIAOs. This work has been supported by the U.S. Department of Energy, Office of Science, Office of Basic Energy Sciences, under Award no. DE-SC0025393.

\appendix

\section{Rotational invariance of the standard matrix elements with GIAOs}\label{rot_proof}

In this Appendix, we will show that the GIAO dressed operators transform correctly under rotation, Eqs. \ref{eq:h_rot}-\ref{eq:pi_rot} and \ref{eq:kI_rot}-\ref{eq:kIL_rot},  which  allowed us to prove angular momentum conservation above. 

\subsection{One-electron matrix elements} 
The first order change of matrix elements for a general one-electron operator $\tilde{\mathcal{O}}^{1}_{\bar{\mu}_{M}\bar{\nu}_{N}}(\bm{R} \bm X) $ is given by: 
\begin{align}
    \zeta(\bm\delta)  = {\tilde{ \mathcal{O}}}^{1}_{\bar{\mu}_{M}\bar{\nu}_{N}}(\bm{R} \bm X)  - \tilde {\mathcal{O}}^{1}_{\mu_{M}\nu_{N}}(\bm X) - O(\bm\delta^2) 
\end{align}
where the rotation matrix is defined as
$\bm{R} = \mathrm{exp}(-\frac{i}{\hbar}\sum_{\alpha}{\bL}^{\alpha}\delta_{\alpha})$, where $\bL^{\alpha}$ is a three dimensional antisymmetric matrix for a rotation around direction $\alpha$; the relevant matrix elements are $L^{\alpha}_{\gamma\beta} = -i\hbar \epsilon_{\alpha\gamma\beta}$. Expanding this operator as a function of the rotation angle $\delta$ gives:
$\bm{R} \approx 1 - \frac{i}{\hbar} \sum_{\alpha}{\bL}^{\alpha}\delta_{\alpha}$. The total rotation of the matrix elements can then  be done in two steps:
\begin{enumerate}
    \item The first step is to rotate the whole molecule, while keeping the orientation of the orbitals fixed. 
    To do so, note that the nuclear gradient of the matrix element can be expanded as:
\begin{align}
\label{eq:1e_R}
    \frac{\partial \tilde{ \mathcal{O}}^{1}_{\mu_{M}\nu_{N}}}{\partial X_{C\gamma}} = \left< \mu_{M} \middle| \frac{\partial \hat{\mathcal{O}}^{1}}{\partial X_{C\gamma}} \middle| \nu_{N}\right>  + \left< \frac{\partial }{\partial X_{C\gamma}} \mu_{M} \middle| {\hat{O}}^{1}\middle| \nu_{N} \right>\delta_{MC} + \left< \mu_{M} \middle|  {\hat{O}}^{1}\middle| \frac{\partial }{\partial X_{C\gamma}}\nu_{N} \right> \delta_{NC}
\end{align}
Above, we have defined (for $\mu$ on $M$ and $\nu$  on $N$), 
    \begin{eqnarray}
        \tilde{ \mathcal{O}}^{1}_{\mu_{M}\nu_{N}} 
        = \left< \mu_M \middle|\exp{\Big(\frac{ie}{\hbar}\bA_{MN} \cdot \br\Big)} \hat{\mathcal{O}}^{1}\middle|\nu_N \right>
        = \left< \mu_M \middle| \hat{\mathcal{O}}^{1}\middle|\nu_N \right>
    \end{eqnarray}
Therefore, all GIAO derivatives are included in the $\tilde{\mathcal{O}}$ term (and not on the basis function $\ket{\nu}$ term). We may evaluate the effect of rotating the molecule:
    \begin{align}
    \label{eq:1e_rot_1}
 \tilde {\mathcal{O}}^{1}_{\mu_{M}\nu_{N}}(\bm{R} \bm X_{0})  - \tilde{ \mathcal{O}}^{1}_{\mu_{M}\nu_{N}}(\bm X_{0})  
     \approx
    \frac{-i}{\hbar}\sum_{C\beta\gamma}\delta_{z}\frac{\partial \tilde{\mathcal{O}}^{1}_{\mu_{M}\nu_{N}}}{\partial X_{C\gamma}}(L^{z}_{\gamma \beta}X_{C\beta}) = \sum_{C\beta\gamma} \delta_{z}\epsilon_{z\gamma\beta} X_{C\beta} \frac{\partial \tilde{\mathcal{O}}^{1}_{\mu_{M}\nu_{N}}}{\partial X_{C\gamma}}
\end{align}

\item  The second step is to rotate the raw basis functions  (without GIAOs) around their atomic centers. The rotated basis functions are of the form: 
    \begin{eqnarray}
        \ket{\bar{\nu}_{N}} \approx \Big(1-\frac{i}{\hbar} \hat{L}^{Nz}_{e} \delta_{z}   \Big)\ket{{\nu}_{N}} 
    \end{eqnarray}
where the $\hat{L}^{Nz}_{e}$ is a rotation operator around atom center $N$ and is given by:
    \begin{eqnarray}
\label{eq:LA_hat}
  \hat{L}^{Nz}_{e}= -i\hbar \sum_{\beta\gamma}\epsilon_{z\beta\gamma}(r_{\beta}-X_{N\beta})\frac{\partial}{\partial  r^{\gamma}} =  \hat{L}_{e}^{z} -  \sum_{\beta\gamma}\epsilon_{z\beta\gamma}X_{N\beta}\hat{p}^{\gamma}_{e}
\end{eqnarray}
If we plug in  Eq. \ref{eq:LA_hat}, so as to rotate  $\ket{\nu_N}$, we can evaluate the contribution to the first order change:
    \begin{align}
    \tilde{\mathcal{O}}^{1}_{{\mu}_{M}\bar{\nu}_{N}}(\bm{R} \bm X_{0}) 
        -\tilde{\mathcal{O}}^{1}_{\mu_{M}\nu_{N}}(\bm{R} \bm X_{0}) \approx
        \frac{-i}{\hbar}\delta_{z}\Bigg( \left<\mu_{M}\middle| \hat{\mathcal{O}}^{1}\middle|\hat{L}^{z}_{e} \nu_{N}\right> - \sum_{\beta\gamma}\epsilon_{z\beta\gamma}X_{N\beta} \left< \mu_{M}\middle|\hat{\mathcal{O}}^{1}\middle| \hat{p}^{\gamma}_{e} \nu_{N}\right> \Bigg)
    \end{align}
Thereafter, we will also rotate  $\bra{{\mu}_{M}}$, leaving
    \begin{eqnarray}
    \label{eq:1e_rot_2}
    & &{\tilde{\mathcal{O}}}^{1}_{\bar{\mu}_{M}\bar{\nu}_{N}}(\bm{R} \bm X_{0}) 
       -{\tilde{\mathcal{O}}}^{1}_{\mu_{M}\nu_{N}}(\bm{R} \bm X_{0})   \\
       \nonumber
       & \approx &  \frac{-i}{\hbar}\delta_{z}\Bigg[
        \left<{\mu_{M}}\middle| [\hat{\mathcal{O}}^{1},\hat{L}^{z}_{e} ]\middle|\nu_{N}\right> + \sum_{\beta\gamma}\epsilon_{z\beta\gamma}\Big(X_{M\beta}\left< \hat{p}^{\gamma}_{e} \mu_{M}\middle|\hat{\mathcal{O}}^{1}\middle|\nu_{N}\right> -X_{N\beta} \left< \mu_{M} \middle| \hat{\mathcal{O}}^{1}\middle|\hat{p}^{\gamma}_{e} \nu_{N}\right> \Big)\Bigg]
    \end{eqnarray}
Finally, we can use the phase condition $(\hat{p}^{\gamma}_{e}  -i\hbar\nabla^{\gamma}_{n}) \ket{ \nu_{N}}=0$, to replace electronic momentum with nuclear momentum. Therefore, the second and third terms on the right hand side of Eq. \ref{eq:1e_rot_2} become 
\begin{eqnarray}
    &&\frac{-i}{\hbar}\sum_{\beta\gamma}\delta_{z}\epsilon_{z\beta\gamma}\Big(X_{M\beta}\left< \hat{p}^{\gamma}_{e} \mu_{M}\middle|\hat{\mathcal{O}}^{1}\middle| \nu_{N}\right> -X_{N\beta} \left< \mu_{M}\middle|\hat{\mathcal{O}}^{1}\middle|\hat{p}^{\gamma}_{e} \nu_{N}\right> \Big) \nonumber \\
&=&-\sum_{\beta\gamma}\delta_{z}\epsilon_{z\beta\gamma}\Bigg(X_{M\beta}\left< \frac{\partial \mu_{M}}{\partial X_{M\gamma}} \middle| \hat{\mathcal{O}}^{1}  \middle| \nu_{N} \right> + X_{N\beta}\left< \mu_{M} \middle| \hat{\mathcal{O}}^{1} \middle| \frac{\partial \nu_{N}}{\partial X_{N\gamma}} \right> \Bigg)\label{eq:1e_rot_2_1}
    \end{eqnarray}
\end{enumerate}
The terms in Eq. \ref{eq:1e_rot_2_1} cancel exactly  with the second and third terms on the right hand side of Eq. \ref{eq:1e_R}. Thus the total rotation after adding the two steps (Eq. \ref{eq:1e_R} and Eq. \ref{eq:1e_rot_2_1}) is: 
\begin{eqnarray}
\zeta(\bm\delta)&=&
\sum_{C\beta\gamma}\delta_{z} \epsilon_{z\beta\gamma} X_{C\beta} \left<{\mu_{M}}\middle| \frac{\partial \hat{\mathcal{O}}^{1}}{\partial X_{C\gamma}}\middle|{\nu_{N}}\right> -\frac{i}{\hbar}\delta_{z} \left<{\mu_{M}}\middle| [\hat{\mathcal{O}}^{1},\hat{L}^{z}_{e} ]\middle|{\nu_{N}}\right>
\label{eq:keep1forlater}
\\
&=& -\frac{i}{\hbar}\delta_{z} \left<{\mu_{M}}\middle| [\hat{\mathcal{O}}^{1},\hat{L}^{z}_{n} +\hat{L}^{z}_{e} ]\middle|{\nu_{N}}\right>\label{eq:1e_Rbar}
\end{eqnarray}
Finally, according to Eq. \ref{eq:1e_Rbar}, we find that, in order for the operator to be rotationally invariant, we require that the operator commutes with the total angular momentum (which was of course expected). 

\subsubsection{One-Electron Core Hamiltonian Matrix Elements}\label{1eHgiao}

For every pair of nuclei, $M$ and $N$, the one-electron Hamiltonian is given by:
\begin{eqnarray}
\tilde{h}_{\mu\nu}(\mathbf{B,X}) &=& \left< \tilde \mu_M \middle| \frac{\hat{\bm{\pi}}_e^2}{2m_e}-\sum_K \frac{Z_K}{\bm{\hat r}_K} \middle| \tilde \nu_N \right> \\&=& \left< \mu_M \middle| \exp{\Big(\frac{ie}{\hbar}\bA_{MN} \cdot \br\Big)} \Big[\frac{\hat{\bm{\pi}}_N^2}{2m_e}-\sum_K \frac{Z_K}{\bm{\hat r}_K}\Big] \middle|  \nu_N \right> \\&=& \left< \mu_M \middle| \hat{h}_{MN}\middle|  \nu_N \right>
\end{eqnarray}
To prove $\hat{h}_{MN}$ is rotationally invariant, we need to show that Eq. \ref{eq:1e_Rbar} holds, i.e., 
\begin{equation}
    [\hat{h}_{MN}, \hat{L}_e^z + \hat{L}_n^z] = 0 \label{eq:comm}
\end{equation}
The total angular momentum commutes with the Hamiltonian and is a good quantum number. Therefore, our task is to show that the GIAO factors also commutes with the Hamiltonian. To that end, we first evaluate the commutator with electronic angular momentum:
\begin{eqnarray}
    \Big[\exp{(\frac{ie}{2\hbar}(\bB\times \bX_{MN})\cdot \br)}, \hat{L}^{z}_{e}\Big] &=& \Big[\exp{(\frac{ie}{2\hbar} B_z({X}^x_{MN}y -X_{MN}^yx))}, \hat{x}\hat{p}_y-\hat{y}\hat{p}_x\Big] \\&=& \hat{x}\Big[\exp{(\frac{ie}{2\hbar} B_z({X}^x_{MN}y -X_{MN}^yx))}, \hp_y\Big]\nonumber\\&&-\hy\Big[\exp{(\frac{ie}{2\hbar} B_z({X}^x_{MN}y -X_{MN}^yx))}, \hp_x\Big] \\
    &=& -\frac{e}{2}B_z\exp{(\frac{ie}{2\hbar}(\bB\times \bX_{MN})\cdot \br)}\Big[\hx X_{MN}^x+ \hy X_{MN}^y  \Big]\nonumber \\\label{eq:LeGIAO}
\end{eqnarray}
Here, we have denoted the components of $\bX_{MN}$ as $(\bX^x_{MN}, \bX^y_{MN}, \bX^z_{MN})$. Second, we evaluate the commutator with nuclear angular momentum:
\begin{eqnarray}
    \Big[\exp{(\frac{ie}{2\hbar}(\bB\times \bX_{MN})\cdot \br)}, \hat{L}^{z}_{n}\Big] &=& \Big[\exp{(\frac{ie}{2\hbar} B_z({X}^x_{MN}y -X_{MN}^yx))}, \sum_I X_IP_{YI} -Y_IP_{XI}\Big]\nonumber \\ \\&=& \sum_I X_I\Big[\exp{(\frac{ie}{2\hbar} B_z({X}^x_{MN}y -X_{MN}^yx))}, P_{YI}\Big]\nonumber\\&&-\sum_I Y_I\Big[\exp{(\frac{ie}{2\hbar} B_z({X}^x_{MN}y -X_{MN}^yx))}, P_{XI}\Big] \\&=&
    -i\hbar\exp{(\frac{ie}{2\hbar} B_z({X}^x_{MN}y -X_{MN}^yx))}\frac{ieB_z}{2\hbar}\nonumber\\
    &&\Big[-xX_M + xX_N - yY_M + yY_N\Big] \\
    &=& - \frac{e}{2}B_z\exp{(\frac{ie}{2\hbar}(\bB\times \bX_{MN})\cdot \br)}\Big[-xX_{MN}^x - yX_{MN}^y \Big]\nonumber \\\label{eq:LnGIAO}
\end{eqnarray}
We see that Eqs. \ref{eq:LeGIAO} and \ref{eq:LnGIAO} cancel each other, which implies that the GIAO factors are rotationally invariant, i.e.,
\begin{eqnarray}
    \Big[\exp{(\frac{ie}{2\hbar}(\bB\times \bX_{MN})\cdot \br)},  \hat{L}^{z}_{e} + \hat{L}^{z}_{n}\Big] &=& 0 \label{eq:giao_commute}
\end{eqnarray}

Thus, we have shown that Eq.  \ref{eq:comm} holds and therefore the one electron core Hamiltonian is rotationally invariant (Eq. \ref{eq:h_rot}).

\subsubsection{Electronic Pseudomomentum}
Next, let us prove that the electronic pseudomomentum like term transforms like a vector under rotation:
\begin{eqnarray}
    \hat{\bm{k}}^I_{\bar{\mu}\bar{\nu}}(\bm{R} \bm X) &=& \bm{R} \hat{\bm{k}}^I_{\mu\nu}(\bm X)\label{eq:kIrot}
\end{eqnarray}
The $\eta$ component of the GIAO-dressed pseudomomentum $\tilde k^{I\eta}_{\mu\nu}$ is given by:
\begin{align}
    \tilde k^{I\eta}_{\mu\nu}
    = \exp{(\frac{ie}{2\hbar}(\bB\times \bX_{MN})\cdot \br)}\Big(\hat{p}_e^\eta+ \frac{e}{2}\sum_{\beta}\epsilon_{\eta z\beta} B_z  (\hat{r}-\frac{X_M+X_N}{2})_\beta-e\sum_{\delta}\epsilon_{\eta z\beta} B_z(\hat{r}-X_I)_\beta\Big)
\end{align}
We have already shown that the GIAO factor $\exp{(\frac{ie}{2\hbar}(\bB\times \bX_{MN})\cdot \br)}$ commutes with the total angular momentum (see Eq. \ref{eq:giao_commute}). Next, we find the commutator with the remaining term:
\begin{align}
   \Bigg[\Big(\hat{p}_e^\eta+ \frac{e}{2}\sum_{\beta}\epsilon_{\eta z\beta} B_z  (\hat{r}-\frac{X_M+X_N}{2})_\beta-e\sum_{\beta}\epsilon_{\eta z\beta} B_z(\hat{r}-X_I)_\beta\Big),\hat{L}_e^z +\hat{L}_n^z\Bigg]\label{eq:Second_commutator}
\end{align}
We can find the commutator of each of the three terms in Eq. \ref{eq:Second_commutator}. For the first term, 
\begin{eqnarray}
    \Big[\hat{p}_e^\eta, \hat{L}_e^z +\hat{L}_n^z\Big] &=& i\hbar \sum_{\beta} \epsilon_{\eta z \beta }\hat{p}_e^\beta \label{eq:kMNIcomm1}
\end{eqnarray}
For the second term in Eq. \ref{eq:Second_commutator}, we find:
\begin{eqnarray}
    & &[\frac{e}{2}\sum_{\beta}\epsilon_{\eta z\beta} B_z  (\hat{r}-\frac{X_M+X_N}{2})_\beta, \hat{L}_e^z +\hat{L}_n^z] \\ 
    & & \; \; \; \; \; \; \; \; = i\hbar\frac{e}{2}\sum_{\beta\gamma}\epsilon_{\eta z\beta}  \epsilon_{\beta z \gamma} B_z \hat{r}_\gamma - i\hbar\frac{e}{2}\sum_{\beta\gamma}\epsilon_{\eta z\beta} \epsilon_{\beta z \gamma} B_z (\frac{X_M+X_N}{2})_\gamma\\
    & & \; \; \; \; \; \; \; \; = -i\hbar\frac{e}{2}\sum_{\gamma}\delta_{\eta \gamma} B_z \hat{r}_\gamma + i\hbar\frac{e}{2}\sum_{\gamma}\delta_{\eta \gamma}B_z (\frac{X_M+X_N}{2})_\gamma\label{eq:kMNIcomm2}
\end{eqnarray}
For the third term in Eq. \ref{eq:Second_commutator}, we find:
\begin{eqnarray}
    \Bigg[\frac{e}{2}\sum_{\beta}\epsilon_{\eta z\beta} B_z  (\hat{r}-X_I)_\beta, \hat{L}_e^z +\hat{L}_n^z\Bigg] 
    = -i\hbar\frac{e}{2}\sum_{\gamma}\delta_{\eta \gamma} B_z \hat{r}_\gamma + i\hbar\frac{e}{2}\sum_{\gamma}\delta_{\eta \gamma}B_z X_{I\gamma}\label{eq:kMNIcomm3}
\end{eqnarray}
Therefore, on combining Eqs. \ref{eq:kMNIcomm1}-\ref{eq:kMNIcomm3} we recover:
\begin{eqnarray}
    \Big[k_{MN}^\eta, \hat{L}_e^z +\hat{L}_n^z\Big] &=& i\hbar \sum_{\beta} \epsilon_{\eta z \beta }\hat{p}_e^\beta -i\hbar\frac{e}{2}\sum_{\gamma}\delta_{\eta \gamma} B_z (\hat{r}-(\frac{X_M+X_N}{2}))_\gamma\\&& -i\hbar\frac{e}{2}\sum_{\gamma}\delta_{\eta \gamma} B_z (\hat{r}-X_I)_\gamma \label{eq:Lkuv}
\end{eqnarray}
The result above must be compared to the first order expansion of the rotated  vector  $\bm{R}\tilde{\bm{k}}_{\mu\nu}$ (where the rotation is around the $z-$ direction):
\begin{eqnarray}
    \bm{R}\tilde{\bm{k}}_{\mu\nu} &=& 
    - \sum_{\beta} \frac{i}{\hbar} L^z_{\eta \beta}\delta_z \tilde k^{I\beta}_{\mu\nu} \\&=&-\sum_\beta\epsilon_{z \eta \beta} \delta_z \tilde k_{\mu\nu}^{I\beta}\\&=&  -\sum_\beta\epsilon_{z \eta \beta} \delta_z \Big(\hat{p}_e^\beta+ \frac{e}{2}\sum_{\gamma}\epsilon_{\beta z\gamma} B_z  (\hat{r}-\frac{X_M+X_N}{2})_\gamma
    +\frac{e}{2}\sum_{\gamma}\epsilon_{\beta z\gamma} B_z  (\hat{r}-X_I)_\gamma\Big)\\&=&  -\sum_\beta\epsilon_{z \eta \beta} \delta_z \hat{p}_e^\beta-\frac{e}{2}\sum_{\beta \gamma}\epsilon_{z \eta \beta} \epsilon_{\beta z\gamma}\delta_z B_z  (\hat{r}-\frac{X_M+X_N}{2})_\gamma
    \nonumber-\frac{e}{2}\sum_{\beta \gamma}\epsilon_{z \eta \beta}\epsilon_{\beta z\gamma}\delta_z B_z  (\hat{r}-X_I)_\gamma\\\\&=&  -\sum_\beta\epsilon_{z \eta \beta} \delta_z \hat{p}_e^\beta-\frac{e}{2}\sum_{\beta \gamma}\delta_{\eta \gamma}\delta_z B_z  (\hat{r}-\frac{X_M+X_N}{2})_\gamma
    -\frac{e}{2}\sum_{\beta \gamma}\delta_{\eta \gamma}\delta_z B_z  (\hat{r}-X_I)_\gamma \label{eq:Rkuv}
\end{eqnarray}
Thus, Eqs. \ref{eq:Lkuv} and \ref{eq:Rkuv} are equivalent and the pseudomomentum vector transforms correctly, i.e., Eq. \ref{eq:kI_rot} (or Eq. \ref{eq:kIrot}) holds. We can similarly prove rotational invariance for the pseudoangular momentum:
\begin{eqnarray}
 (\hbr\times \hat{\bm{k}}^I)_{\bar{\mu}\bar{\nu}}(\bm{R} \bm X) &=& \bm{R} (\hbr\times \hat{\bm{k}}^I)_{\mu\nu}(\bm X)   
\end{eqnarray}

\subsection{Two-electron matrix elements}
The two electron operator Coulomb operator in a GIAO basis is given by:
\begin{eqnarray}
    \tilde g_{\mu\nu \lambda \sigma}(\mathbf{B,X}) &=& \left< \mu_M \nu_N \middle| \exp{\Big(\frac{ie}{2\hbar} \bB\cdot (\mathbf{X}_{MP} \times \mathbf{ r_1} + \mathbf{X}_{NQ} \times \mathbf{ r_2})\Big)}\bm{\hat{r}}_{12}^{-1}\middle| \lambda_P \sigma_Q \right> \\
    &=& \left< \mu_M \nu_N \middle| \hat{g}_{MNPQ}\middle| \lambda_P \sigma_Q \right>
\end{eqnarray}
When rotating any  two-electron matrix element $\hat{\mathcal{O}}^{2}(\bm{R} \bm X) $ around the $z-$axis, the analogue to Eq. \ref{eq:keep1forlater} is given by: 
\begin{eqnarray}
   \zeta(\bm \delta)&=& \sum_{Q\beta\gamma} \delta_{z}\epsilon_{\beta\gamma z} X_{Q\beta} \left<\mu_{M}\nu_{N} \middle|\frac{\partial \hat{\mathcal{O}}^{2}}{\partial X_{Q\gamma}}\middle |\lambda_{P}\sigma_{Q}\right>-  \frac{i}{\hbar}\delta_{z}
        \left<\mu_{M}\nu_{N} \middle| [\hat{\mathcal{O}}^{2},\hat{L}^{z}_{e} ]\middle|\lambda_{P}\sigma_{Q}\right> \nonumber \\
        &=& -  \frac{i}{\hbar} \delta_{z}
        \left<\mu_{M}\nu_{N} \middle| [{\hat{{\mathcal{O}}}}^{2},\hat{L}^{z}_{n}+\hat{L}^{z}_{e} ]\middle|\lambda_{P}\sigma_{Q}\right>\label{eq:2rot}
\end{eqnarray}
An explicit derivation of Eq. \ref{eq:2rot} is given in Ref. \citenum{coraline_pssh}. Obviously, the $\bm{\hat{r}}_{12}^{-1}$ operator commutes with angular momentum so that the bare two-electron matrix elements are rotationally invariant. Now, 
in Sec. \ref{1eHgiao}, we have shown that the GIAO phases operators also commute with the total angular momentum. Thus, the GIAO dressed two-electron matrix elements  are also rotationally invariant.

\section{Angular Momentum and Pseudomomentum Conservation in GIAO basis}\label{am-pm_conserve}

For the sake of completeness, and the convenience of the reader who may not have time to go through Paper I, we will here give a tedious but thorough proof of pseudomomentum and angular momentum conservation (Secs. \ref{pmconservedefn} and \ref{amconservedefn} above) for the phase-space Hamiltonian with a GIAO basis -- but now without reference to Paper I.

\subsection{Pseudomomentum Conservation}
For a collection of particles (with charges $Z_Ie$) in a magnetic field, the total conserved quantity is the sum of the pseudomomentum of the nuclei and the GIAO-dressed pseudomomentum like term for the electrons:
\begin{eqnarray}
\label{eq:pseudomom}
   K^\alpha_{mol} &=& \sum_I {P_{I\alpha}}+ \frac{1}{2}\sum_I q^{\textit{eff}}_I(\bX)(\bB\times \bX_I)_{\alpha}-i\hbar \sum_{I\mu\nu} \tilde{D}_{\nu\mu}\tilde{{\Gamma}}_{\mu\nu}^{I\alpha} + \sum_{\mu \nu} \tilde{D}_{\nu \mu} \tilde {k}^{I\alpha}_{\mu\nu} \\ &=& \sum_I {P_{I\alpha}}+ \frac{1}{2}\sum_I q^{\textit{eff}}_I(\bX)(\bB\times \bX_I)_{\alpha}
\end{eqnarray}
In Eq. \ref{eq:pseudomom}, the last two terms cancel using condition 1 in Eq. \ref{eq:Gamma_uv1}. To prove that the pseudomomentum is a constant of motion we can evaluate its time derivative, using the equations of motions in Eqs. \ref{eq:xdot}-\ref{eq:pdot}:
\begin{eqnarray}
   \frac{d  K^\alpha_{mol}}{dt} &=& \sum_I \dot P_{I\alpha} + \frac{1}{2}\sum_Iq_I^{\textit{eff}}({\bX})(\bB \times \dot \bX_I)_{\alpha} \\ 
   &=&-\sum_{IJ\beta} \frac{\Pi^{\textit{eff}}_{J\beta}}{M_J}\Big(- \frac{1}{2} \frac{\partial q^{\textit{eff}}_J(\bX)}{\partial X_{I\alpha}}(\bB\times \bX_J)_{\beta}- \frac{1}{2} q^{\textit{eff}}_J(\bX)\frac{\partial (\bB\times \bX_J)_{\beta}}{\partial X_{I\alpha}}\Big)\nonumber\\
   &&+i\hbar \sum_{IJ\beta \mu\nu}   \frac{\tilde D_{\nu\mu}}{M_J}\Bigg(- \frac{1}{2} \frac{\partial q^{\textit{eff}}_J(\bX)}{\partial X_{I\alpha}} (\bB\times \bX_J)_{\beta} - \frac{1}{2} q^{\textit{eff}}_J(\bX)\frac{\partial (\bB\times \bX_J)_{\beta} }{\partial X_{I\alpha}}\Bigg) \tilde{\Gamma}^{J\beta}_{\mu\nu}\nonumber\\&&-\sum_I\left(\frac{\partial V_{PS}}{\partial X_{I \alpha}}\right)_{\bm \Pi^\textit{eff}}+ \frac{1}{2}\sum_Iq_I^{\textit{eff}}({\bX})(\bB \times \dot \bX_I)_{\alpha}
\end{eqnarray}
Since we are using GIAOs, the total phase-space potential energy term is translationally invariant as shown in Eq. \ref{eq:transEgiao}. Moreover, $q^{\textit{eff}}_I$ is also translationally invariant, i.e., $\sum_{IJ}\frac{\partial q^{\textit{eff}}_J(\bX)}{\partial X_{I}}=0$. Thus, finally the equations can be written as:
\begin{eqnarray}
\frac{d  K^\alpha_{mol}}{dt} &=&-\sum_{IJ\beta} \frac{\Pi^{\textit{eff}}_{J\beta}}{M_J}\Big(- \frac{1}{2} q^{\textit{eff}}_J(\bX)\frac{\partial (\bB\times \bX_J)_{\beta}}{\partial X_{I\alpha}}\Big)_{\bm P} \nonumber\\
   &&+i\hbar \sum_{IJ\beta \mu\nu}  \frac{\tilde D_{\nu\mu}}{M_J}\Big(- \frac{1}{2} q^{\textit{eff}}_J(\bX)\frac{\partial (\bB\times \bX_J)_{\beta} }{\partial X_{I\alpha}}\Big)_{\bm P} \tilde{\Gamma}^{J\beta}_{\mu\nu}\nonumber\\&&+ \frac{1}{2}\sum_Iq_I^{\textit{eff}}({\bX})(\bB \times \dot \bX_I)_{\alpha}
\end{eqnarray}
Further expanding $\dot{X}_{I\alpha}$  (from Eq. \ref{eq:xdot}) and taking the gradient of the vector potential gives:
\begin{eqnarray}
   \frac{d K^\alpha_{mol}}{dt} &=&\frac{1}{2}\sum_{IJ\beta}  q^{\textit{eff}}_J(\bX) \epsilon_{\beta z \alpha}B_z\frac{\Pi^{\textit{eff}}_{J\beta}}{M_J}\delta_{IJ}-\frac{i\hbar}{2}\sum_{I J\beta\mu\nu}\tilde{D}_{\nu\mu}q^{\textit{eff}}_J(\bX)\epsilon_{\beta z \alpha}B_z  \frac{ \tilde{\Gamma}^{J\beta}_{\mu\nu}}{M_{J}} \delta_{IJ}\nonumber\\
   &&+ \frac{1}{2}\sum_{I\beta}q^{\textit{eff}}_I(\bX) \epsilon_{\alpha z \beta}B_z\frac{\Pi^{\textit{eff}}_{I\beta}}{M_{I}} - \frac{i\hbar}{2}\sum_{I\beta \nu\mu} \tilde{D}_{\nu\mu}q^{\textit{eff}}_I(\bX) \epsilon_{\alpha z \beta}B_z \frac{ \hat{\Gamma}^{I\beta}_{\mu\nu}}{M_{I}} \\
   &=& 0\label{eq:pseudoconsv}
\end{eqnarray}
Therefore, the pseudomomentum in all $\alpha$ directions is conserved in a uniform magnetic field. 

\subsection{Angular momentum conservation}\label{ang_conserve}
In the presence of the magnetic field, the conserved quantity is the canonical angular momentum of the nuclei around a gauge $\bG$ plus the angular momentum dressed with GIAOs for the electrons around the same gauge origin $\bG$: 
\begin{eqnarray}
    L_{mol}^z &=& \sum_I \Bigg((\bX_I-\bG) \times \Big( \bm{\Pi}^{\textit{eff}}_{I} - i\hbar \sum_{\mu\nu}\tilde{D}_{\nu\mu}\bm{\tilde{\Gamma}}^{I}_{\mu\nu}+\frac{q_I^{\textit{eff}}(\bX)}{2}(\bB\times (\bX_I-\bG))\Big)\Bigg)_z \nonumber\\&&+ \sum_{I\mu\nu}  \hat{\Theta}_I\tilde D_{\nu\mu}{(( \hbr- \bG)\times \tilde{\bm{k}}^I_{\mu\nu})}_{z}
\end{eqnarray}
To show that the total angular momentum is conserved, we evaluate its time derivative:
\begin{eqnarray}
\frac{d}{dt} L^{z}_{mol} &=& \frac{d}{dt}\Big[\sum_{I\beta\gamma} \epsilon_{z\beta\gamma}(X_{I\beta}-G_{\beta})M_{I}\Big( \frac{ \Pi^{\textit{eff}}_{I\gamma}}{M_{I}} - i\hbar\sum_{\mu\nu}\tilde D_{\nu\mu}\frac{ \tilde{\Gamma}^{I\gamma}_{\mu\nu}}{M_{I}} \Big) \nonumber\\
&&+ \sum_{I\beta\gamma} \epsilon_{z\beta\gamma}(X_{I\beta}-G_{\beta})\frac{q_I^{\textit{eff}}(\bX)}{2}(\bB\times (\bX_I-\bG))_\gamma \nonumber\\
&&+ \sum_{I\mu\nu} \tilde{D}_{\nu\mu} \Big(( \tilde \br- \bG)\times \frac{1}{2} ( \hat{\Theta}_I \tilde{\bm{k}}_{I} +  \tilde{\bm{k}}_{I} \hat{\Theta}_I ) \Big)^z_{\mu\nu} \Big]
 \label{eq:lst} 
\end{eqnarray}
To proceed further, we use Eq. \ref{eq:Gamma_uv3} to cancel out the terms involving $\tilde \Gamma_{\mu\nu}$ and electronic pseudomomentum $\tilde k^I_{\mu\nu}$ in Eq. \ref{eq:lst}. 
\begin{eqnarray}
\frac{d}{dt} L^{z}_{mol} 
&=& \frac{d}{dt} \Big[\sum_{I\beta\gamma} \epsilon_{z\beta\gamma}(X_{I\beta}-G_{\beta})(P_{I\gamma}-\frac{q_I^{\textit{eff}}(\bX)}{2}(\bB\times (\bX_I-\bG))_\gamma)\nonumber\\&&+ \sum_{I\beta\gamma} \epsilon_{z\beta\gamma}(X_{I\beta}-G_{\beta})\frac{q_I^{\textit{eff}}(\bX)}{2}(\bB\times (\bX_I-\bG))_\gamma \Big] \\
&= & \frac{d}{dt}\Big[\sum_{I\beta\gamma} \epsilon_{z\beta\gamma}(X_{I\beta}-G_{\beta})P_{I\gamma}\Big]\\
&=&  \sum_{I\beta\gamma}\epsilon_{z\beta\gamma}\Big(\dot X_{I\beta} P_{I\gamma} + (X_{I\beta}-G_{\beta})\dot P_{I\gamma}\Big) 
\\
 &=&  -\sum_{I\beta\gamma}\epsilon_{z\beta\gamma}\Big(P_{I\beta} \left(\frac{\partial E_{PS}}{\partial P_{I\gamma}}\right)_{\bm X}+ (X_{I\beta}-G_\beta) \left(\frac{\partial E_{PS}}{\partial X_{I\gamma}}\right)_{\bm P}\Big)\label{eq:rot_inv}
\end{eqnarray}
Because Sec. \ref{gamma_rot} demonstrates that the phase-space energy is rotationally invariant around the $z-$axis, it follows that the total angular momentum is conserved. 

\bibliography{Ref}
\end{document}